\PassOptionsToPackage{unicode}{hyperref}
\PassOptionsToPackage{hyphens}{url}
\documentclass[
  11pt,
]{article}
\usepackage{amsmath,amssymb}
\usepackage{setspace}
\usepackage{iftex}
\ifPDFTeX
  \usepackage[T1]{fontenc}
  \usepackage[utf8]{inputenc}
  \usepackage{textcomp} 
\else 
  \usepackage{unicode-math} 
  \defaultfontfeatures{Scale=MatchLowercase}
  \defaultfontfeatures[\rmfamily]{Ligatures=TeX,Scale=1}
\fi
\usepackage{lmodern}
\ifPDFTeX\else
\fi
\IfFileExists{upquote.sty}{\usepackage{upquote}}{}
\IfFileExists{microtype.sty}{
  \usepackage[]{microtype}
  \UseMicrotypeSet[protrusion]{basicmath} 
}{}
\makeatletter
\@ifundefined{KOMAClassName}{
  \IfFileExists{parskip.sty}{%
    \usepackage{parskip}
  }{
    \setlength{\parindent}{0pt}
    \setlength{\parskip}{6pt plus 2pt minus 1pt}}
}{
  \KOMAoptions{parskip=half}}
\makeatother
\usepackage{xcolor}
\usepackage[margin=1in]{geometry}
\usepackage{color}
\usepackage{fancyvrb}

\DefineVerbatimEnvironment{Highlighting}{Verbatim}{commandchars=\\\{\}}
\usepackage{framed}
\definecolor{shadecolor}{RGB}{248,248,248}
\newenvironment{Shaded}{\begin{snugshade}}{\end{snugshade}}

\newcommand{\AttributeTok}[1]{\textcolor[rgb]{0.13,0.29,0.53}{#1}}

\newcommand{\CommentTok}[1]{\textcolor[rgb]{0.56,0.35,0.01}{\textit{#1}}}

\newcommand{\DecValTok}[1]{\textcolor[rgb]{0.00,0.00,0.81}{#1}}

\newcommand{\FloatTok}[1]{\textcolor[rgb]{0.00,0.00,0.81}{#1}}
\newcommand{\FunctionTok}[1]{\textcolor[rgb]{0.13,0.29,0.53}{\textbf{#1}}}

\newcommand{\NormalTok}[1]{#1}

\newcommand{\OtherTok}[1]{\textcolor[rgb]{0.56,0.35,0.01}{#1}}

\newcommand{\SpecialCharTok}[1]{\textcolor[rgb]{0.81,0.36,0.00}{\textbf{#1}}}

\newcommand{\StringTok}[1]{\textcolor[rgb]{0.31,0.60,0.02}{#1}}

\usepackage{longtable,booktabs,array}
\usepackage{calc} 
\usepackage{etoolbox}
\makeatletter
\patchcmd\longtable{\par}{\if@noskipsec\mbox{}\fi\par}{}{}
\makeatother
\IfFileExists{footnotehyper.sty}{\usepackage{footnotehyper}}{\usepackage{footnote}}
\makesavenoteenv{longtable}
\usepackage{graphicx}
\makeatletter
\def\maxwidth{\ifdim\Gin@nat@width>\linewidth\linewidth\else\Gin@nat@width\fi}
\def\maxheight{\ifdim\Gin@nat@height>\textheight\textheight\else\Gin@nat@height\fi}
\makeatother
\setkeys{Gin}{width=\maxwidth,height=\maxheight,keepaspectratio}
\makeatletter
\def\fps@figure{htbp}
\makeatother
\providecommand{\tightlist}{%
  \setlength{\itemsep}{0pt}\setlength{\parskip}{0pt}}
\newlength{\cslhangindent}
\newlength{\csllabelwidth}
\newlength{\cslentryspacingunit} 
\newenvironment{CSLReferences}[2] 
 {
  \setlength{\parindent}{0pt}
  \ifodd #1
  \let\oldpar\par
  \def\par{\hangindent=\cslhangindent\oldpar}
  \fi
  \setlength{\parskip}{#2\cslentryspacingunit}
 }%
 {}
\usepackage{calc}

\usepackage{txfonts}
\usepackage{lineno}
\usepackage{caption}
\usepackage{bm}
\usepackage{booktabs}

\usepackage{flafter}
\ifLuaTeX
  \usepackage{selnolig}  
\fi
\IfFileExists{bookmark.sty}{\usepackage{bookmark}}{\usepackage{hyperref}}
\IfFileExists{xurl.sty}{\usepackage{xurl}}{} 
\hypersetup{
  pdftitle={Extended molt phenology models improve inferences about molt duration and timing},
  pdfauthor={Philipp H. Boersch-Supan1*, Hugh J. Hanmer1 \& Robert A. Robinson1},
  hidelinks,
  pdfcreator={LaTeX via pandoc}}

\title{Extended molt phenology models improve inferences about molt duration and timing}
\author{Philipp H. Boersch-Supan\textsuperscript{1}*, Hugh J. Hanmer\textsuperscript{1} \& Robert A. Robinson\textsuperscript{1}}
\date{04 December 2023}

\begin{document}
\maketitle

\setstretch{1.5}
\small

\textsuperscript{1} British Trust for Ornithology, The Nunnery, Thetford IP24 2PU, United Kingdom;\\
\texttt{*} Corresponding author: \href{mailto:pboesu@gmail.com}{\nolinkurl{pboesu@gmail.com}}
\normalsize

\hypertarget{acknowledgements}{%
\section*{Acknowledgements}\label{acknowledgements}}
\addcontentsline{toc}{section}{Acknowledgements}

We thank the thousands of fieldworkers and volunteers who contribute records to the British and Irish Ringing Scheme and other schemes collecting molt data worldwide, and in particular Hugh Insley for providing the Siskin dataset.
Simulation studies were conducted on JASMIN, the UK's collaborative data analysis environment (\url{https://jasmin.ac.uk}).

\textbf{Funding statement:} No external funding was provided for this study.

\textbf{Ethics statement:} This research was conducted in compliance with the British Trust for Ornithology's Code of Good Scientific Practice.

\textbf{Conflict of interest statement:} The authors declare no competing interests.

\textbf{Author contributions:} PHBS and RAR conceived the study; PHBS led model development, analysis, and writing with input from all authors; HJH contributed to model testing; all authors contributed critically to the manuscript and gave final approval for publication.

\textbf{Data availability:} Data and code for the simulation and case studies are archived on Zenodo: \url{https://doi.org/10.5281/zenodo.7643922}

\newpage

\

\textbf{Abstract}\\
Molt is an essential life-history event in birds and many mammals, as maintenance of feathers and fur is critical for survival.
Despite this molt remains an understudied life-history event.
Non-standard statistical techniques are required to estimate the phenology of molt from observations of plumage or pelage state, and existing molt phenology models have strict sampling requirements which can be difficult to meet under real-world conditions.
We present an extended modelling framework that can accommodate features of real-world molt datasets, such re-encounters of individuals, misclassified molt states, and/or molt state-dependent sampling bias.
We demonstrate that such features can lead to biased inferences when using existing molt phenology models, and show that our model extensions can improve inferences about molt phenology under a wide range of sampling conditions.
We hope that our novel modelling framework removes barriers for modelling molt phenology data from real-world datasets and thereby further facilitates the uptake of appropriate statistical methods for such data.
Although we focus on molt, the modelling framework is applicable to other phenological processes which can be recorded using either ordered categories or approximately linear progress scores.

\textbf{Keywords:} Molt; phenology; plumage; pelage; Bayesian inference; individual heterogeneity; mark-recapture;

\textbf{Lay summary}

\begin{itemize}
\tightlist
\item
  Molt is an essential but understudied life-history event in birds and many mammals.
\item
  The analysis of molt data requires non-standard statistical techniques and existing molt phenology models have strict sampling requirements which can be difficult to meet.
\item
  We present new statistical models for molt timing and duration.
\item
  We demonstrate that these models improve inferences about molt duration and timing under a wide range of realistic sampling conditions.
\end{itemize}

\hypertarget{introduction}{%
\section*{Introduction}\label{introduction}}
\addcontentsline{toc}{section}{Introduction}

All birds and most mammals rely on feathers and fur, respectively, to deliver critical functions such as thermoregulation (Wolf and Walsberg 2000, Dawson et al. 2014), social signaling and crypsis (Caro 2005, McQueen et al. 2019), or flight (Matloff et al. 2020).
Plumage and pelage are non-living keratinous structures subject to continuous wear and therefore must be replaced regularly through shedding and regrowth, a replacement process known as molt (Ling 1972, Newton 2009).
Molt is an essential life-history event, which - unlike breeding - cannot be skipped, as the maintenance of pelage or plumage is necessary for survival (Ling 1970, Jenni and Winkler 2020).
Birds and mammals have evolved a bewildering array of strategies to balance the molt process with demands imposed by breeding and migration (Beltran et al. 2018).
Despite this, many aspects of the biology of molt, including environmental and behavioural drivers of its timing, extent and rate, remain poorly known (Bridge 2011, Marra et al. 2015), even though they may hold crucial clues to understanding animals' responses to a changing environment (Tomotani et al. 2018, Zimova et al. 2020b, Kock et al. 2021, Hanmer et al. 2022).

Phenology, the timing of seasonal biological phenomena, is a key aspect of plant and animal life.
It defines the timing and duration of life-cycle events and thereby determines the ability of organisms to capture seasonally variable resources (Chuine and Régnière 2017).
Phenological analyses often focus on the timing of particular life-history events, such as the dates of egg laying (Shutt et al. 2019) or parturition (Moyes et al. 2011).
However, for many biological phenomena, including molt, exact dates of particular events are more difficult to observe than the state of the system itself.
In most studies of free-living animal populations the initiation, progression, and completion of molt cannot be observed fully in individuals.
Rather, observations generally consist of snapshots of individuals that have old plumage or pelage, are in some stage of active molt, or have new plumage or pelage.
Transition dates and the duration of molt in the population then have to be inferred from such data and this generally precludes the use of simple linear regression methods.

As an additional complication, the observation of molt in the field can vary both in terms of which stages of the population can be sampled and how molt progress is recorded.
For birds, two main types of recording systems exist for molt status (e.g. Dolnik and Gavrilov 1974, Ginn and Melville 1983): a simpler, qualitative type records molt status as a categorical variable (not started, in progress/active molt, completed), whereas a more quantitative type records the progression of individual feathers or feather tracts for birds in the active molt category.
The latter is commonly scored in the field using a six-point scale where 0 is an old unmolted feather and scores 1 to 5 represent stages of feather growth from the shedding of the old feather to a fully grown new feather (Ginn and Melville 1983).
The sum of the primary feather scores for one wing (a scale of 0-45 or 50 for most passerine birds, which have 9-10 fully developed primary feathers) gives a quantitative measure of overall molt progression.
Where feathers differ in size, raw molt scores do not necessarily increase linearly with time, which may be ameliorated by converting raw scores into the proportion of new primary feather mass grown (PFMG), using feather specific masses (Underhill and Joubert 1995).
In mammals molt progress is typically described as the proportion of the
molted body surface, often recorded under field conditions in a simplified manner as ordered categories that may or may not represent equal intervals of molt progress (e.g. Watson 1963, Beltran et al. 2019, Zimova et al. 2020a, Kock et al. 2021).

Underhill and Zucchini (1988) proposed a general modelling framework for both categorical and molt score data, applicable to species with a continuous molt strategy.
Maximum likelihood inference for these models is implemented in the \emph{moult} package for R (Erni et al. 2013).
For categorical molt data, an approach based on the probit model was subsequently suggested by Rothery and Newton (2002), which can be implemented in general purpose GLM software.
Both approaches for categorical molt data are special cases of ordinal regression models (Bürkner and Vuorre 2019), and are closely related to phenological models developed for specific applications across various ecological disciplines (e.g. Dennis et al. 1986, Candy 1991, Zimova et al. 2020b, Boersch-Supan 2021b), whereas the model for continuous score data has parallels to censored regression models such as the tobit model (Twisk and Rijmen 2009).

Underhill-Zucchini molt models assume that any variation in the molt duration can be modelled using covariates, that the times of molt onset follow a specified probability distribution, and, importantly, that individuals caught on each sampling occasion are a random sample of the modelled population.
The latter implies that sampling probabilities need to be independent of molt stage and homogeneous for all individuals within each molt stage.
This homogeneous sampling assumption is critical, as violations of it can lead to biased parameter estimates, as demonstrated by Bonnevie (2010a) and in the simulation studies below, and ultimately biased ecological conclusions.

However, the assumptions underlying the Underhill-Zucchini molt models, in particular with respect to random sampling across molt stages, can be difficult to meet with real-world molt data.
This is a consequence of the complexities of molt itself as a biological process, as well as the vagaries of obtaining molt data in the field.
In particular, uneven sampling across molt state is common in real-world molt data sets.
This can have biological reasons, as individuals in peak molt are often restricted in their ability to move and may therefore have a different detectability than non-molting individuals.
For example, molting birds may be more cryptic than non-molting birds (Newton 1966, Haukioja 1971), while molting pinnipeds may be easier to observe at a haul-out than mobile non-molting individuals (Watts 1996).
Alternatively, uneven sampling can result from the data collection process, for example when molt data arises from opportunistic ringing schemes with irregular sampling schedules.
As a possible result of this, ad hoc statistical approaches have remained common in the analysis of molt data (e.g. Beltran et al. 2019, Kock et al. 2021, Mumme et al. 2021b).
The lack of formal molt models which can simultaneously estimate the full phenological distribution of molt and accommodate imperfect real-world data limits the inferences that can be drawn about physiological, environmental, and behavioural drivers of the timing, extent and rate of molt and ultimately the drivers of animals' annual cycles.

We present extensions to Underhill-Zucchini molt phenology models by using a Bayesian inference framework for parameter estimation that facilitates the analysis of real-world molt data and improves inferences about molt duration and timing under a wide range of realistic sampling conditions.
Although devised for avian primary feather molt, this modelling framework is applicable to all phenological processes which can be recorded using either ordered categories or approximately linear progress scores between well-defined start and end states.

\hypertarget{methods}{%
\section*{Methods}\label{methods}}
\addcontentsline{toc}{section}{Methods}

\hypertarget{molt-model-types}{%
\subsection*{Molt model types}\label{molt-model-types}}
\addcontentsline{toc}{subsection}{Molt model types}

Five Underhill-Zucchini model types are distinguished depending on the molt data type (categorical or score) and the stages of molt that are sampled across the population (Underhill and Zucchini 1988, Underhill et al. 1990):
The type 1 model is for purely categorical observations (molt not started, molt in progress/active molt, molt completed), whereas model types 2-5 require molt scores for all actively molting individuals.
Type 1 (categorical) and type 2 models further require that pre-molt, molt and post-molt individuals are all sampled.
The type 3 model requires samples of active molt individuals only, while
type 4 and 5 models were motivated by species that molt immediately before or after migration, and are therefore not observable across all three molt categories (Tab. \ref{tab:data-type-table}).

\begin{table}

\caption{\label{tab:data-type-table}Sampling situations and data types distinguished in the Underhill-Zucchini molt modelling framework.}
\centering
\begin{tabular}[t]{rlll}
\toprule
Model Type & Pre-molt & Molt & Post-molt\\
\midrule
1 & categorical/0 & categorical & categorical/1\\
2 & categorical/0 & score (0,1) & categorical/1\\
3 & not observed & score (0,1) & not observed\\
4 & not observed & score (0,1) & categorical/1\\
5 & categorical/0 & score (0,1) & not observed\\
\addlinespace
12 & categorical/0 & mix of categorical and score (0,1) & categorical/1\\
\bottomrule
\end{tabular}
\end{table}

\hypertarget{basic-likelihood-functions-for-different-data-types}{%
\subsection*{Basic likelihood functions for different data types}\label{basic-likelihood-functions-for-different-data-types}}
\addcontentsline{toc}{subsection}{Basic likelihood functions for different data types}

Data likelihoods for Underhill-Zucchini models follow a modular structure.
Following the notation of Underhill and Zucchini (1988), samples consist of \(I\) pre-molt individuals, \(J\) individuals in active molt, and \(K\) post-molt individuals.
Individuals in each category are observed on days \(t = t_1,\ldots,t_I\); \(u = u_1,\ldots,u_J\); \(v = v_1,\ldots,v_K\), respectively.
Molt scores for actively molting individuals, where available, are encoded as \(y = y_1,\ldots,y_J\).

For a given date \(t\) each molt state has a probability of occurrence

\begin{align}
P(t) &= \Pr\{Y(t)=0\}&= &1-F_T(t) &\label{eq:Pt}\\ 
Q(t) &= \Pr\{0<Y(t)<1\}& =&F_T(t)-F_T(t-\tau) &\label{eq:Qt}\\
R(t) &= \Pr\{Y(t)=1\}&= &F_T(t-\tau) &\label{eq:Rt}
\end{align}

Where \(F_T\) is the cumulative probability function of the molt initiation dates of individuals in the population, and \(\tau\) is the mean duration of molt in the population.
Note that in the notation of Underhill and Zucchini (1988) \(t\) is doubly defined. It is both a generic variable of time in the model derivation, and when indexed, denotes the sample dates of pre-molt birds in the data likelihoods
Further, assuming a linear progression of the molt indices over time, the probability density of a particular molt score at time \(t\) is

\begin{equation}
q(t,y)=f_Y(t)(y)=\tau f_T(t-y\tau),\quad0 < y < 1,\quad\label{eq:qty}
\end{equation}

In existing implementations the unobserved molt start dates of individuals in the study population are assumed to follow a normal distribution with mean molt start date \(\mu\) and population standard deviation of molt start dates \(\sigma\), such that

\begin{equation}
F_T(t)=\Phi\left(\frac{t-\mu}{\sigma}\right)\quad\label{eq:FT}
\end{equation}

where \(\Phi\) is the standard normal distribution function and

\begin{equation}
f_T(t) = \phi(t) = \frac{1}{\sqrt{2\pi}}\exp\frac{-t^2}{2}.\quad\label{eq:fT}
\end{equation}

It is further assumed that the likelihood has \(p\) parameters \(\boldsymbol{\theta} = (\theta_1, \theta_2, \ldots, \theta_p)\).
If there are no covariate data, then \(p=3\) and \(\boldsymbol{\theta}=(\mu, \tau, \sigma)\).
Otherwise the effect of covariates on the molt parameters can be modelled using linear predictors on \(\mu\) and \(\tau\), and log-linear predictors on \(\sigma\) (Erni et al. (2013)), i.e.~

\begin{align}
\mu &= \mu_0 + \mathbf{X}_\mu\boldsymbol{\beta}_\mu &\label{eq:mu-linpred}\\
\tau &= \tau_0 + \mathbf{X}_\tau\boldsymbol{\beta}_\tau &\label{eq:tau-linpred}\\
\sigma &= \exp(\sigma_0 + \mathbf{X}_\sigma\boldsymbol{\beta}_\sigma) &\label{eq:sigma-linpred}
\end{align}

and consequently \(\boldsymbol{\theta}=(\mu_0, \boldsymbol{\beta}_\mu, \tau_0, \boldsymbol{\beta}_\tau, \sigma_0, \boldsymbol{\beta}_\sigma)\).

Full likelihood expressions for model types 1-5, which represent typical combinations of available molt records are given in Supplementary Materials \ref{sec:likelihoods-appendix}.
Molt parameters for model types 1-5 can be estimated using a maximum likelihood approach, implemented in R package \emph{moult} (Erni et al. 2013).
We here also introduce a Bayesian estimation approach using fast Hamiltonian Monte Carlo samplers (Carpenter et al. 2017), which is implemented in R package \emph{moultmcmc} (Boersch-Supan et al. 2023a).
Further details of the software implementation and installation instructions are provided in Supplementary Materials \ref{sec:software-appendix}.
A joint likelihood for data of type 1 and 2 is straightforward to derive by combining the type 1 and 2 data likelihoods, as outlined in Underhill and Zucchini (1988), and this integrated Type 12 model (T12; Supplementary Materials \ref{sec:likelihoods-appendix}) is also implemented in \emph{moultmcmc}.

\hypertarget{extensions-to-the-underhill-zucchini-framework}{%
\subsection*{Extensions to the Underhill-Zucchini framework}\label{extensions-to-the-underhill-zucchini-framework}}
\addcontentsline{toc}{subsection}{Extensions to the Underhill-Zucchini framework}

We propose extensions to the Underhill-Zucchini modelling framework to address two features of real-world molt data: the misclassification of non-molting individuals, and the presence of molt status-dependent sampling bias.
These methods are implemented in the R package \emph{moultmcmc} (Boersch-Supan et al. 2023a)

\hypertarget{misclassification-of-non-molting-individuals-the-lumped-molt-model}{%
\subsubsection*{Misclassification of non-molting individuals: the lumped molt model}\label{misclassification-of-non-molting-individuals-the-lumped-molt-model}}
\addcontentsline{toc}{subsubsection}{Misclassification of non-molting individuals: the lumped molt model}

In birds, active primary molt is generally unambiguously identifiable on a specimen, but the distinction between the two non-molting categories can be more challenging, both conceptually and practically.
The transition from old plumage to active primary molt, and the transition from active molt to new plumage are reasonably discrete, marked by the shedding of the first feather, and the completion of growth on the last, respectively.
These transitions are conceptually straightforward, and the natural order of the categories across the molt process motivate the framing of these models as ordered categorical regression models.
However, this conceptualisation ignores feather or fur wear between successive molt seasons, i.e.~the gradual transition from new to old plumage or pelage following the completion of a molt cycle.
This can make the assignment of non-molting individuals to pre-molt and post-molt categories ambiguous, both because wear can be difficult to assess, leading to misclassified data, and - when the molt season is long - individuals in worn and unworn stages can co-exist in the population (e.g. Craig et al. 2014), which is incongruous with the notion of linear time implied by the ordered categorical framework.

This classification problem can be sidestepped by distinguishing only two categories of individuals: molting and non-molting individuals.
The probability of occurrence for the former is the same as in the standard model (i.e.~\(Q(t)\), Eq. \eqref{eq:Qt} and/or \(q(t,y)\), Eq. \eqref{eq:qty}).
The probability of the latter then follows from Eq. \eqref{eq:Pt} - \eqref{eq:Rt} as
\begin{equation}
PR(t) = 1 - Q(t) = P(t) + R(t)\quad\label{eq:PRt}
\end{equation}
Because the two non-molting categories are treated as indistinguishable, we refer to this model variant as the \emph{lumped} model below, and denote it with the letter L in abbreviations, e.g.~the lumped type 2 model (T2L).

\hypertarget{repeated-measures-data-the-recaptures-model}{%
\subsubsection*{Repeated measures data: the recaptures model}\label{repeated-measures-data-the-recaptures-model}}
\addcontentsline{toc}{subsubsection}{Repeated measures data: the recaptures model}

A large amount of avian molt data is collected in conjunction with bird ringing (Ginn and Melville 1983, Rose et al. 2020), and as a result mark-recapture information may be available for the individuals for which molt status is recorded.
Similarly, many demographic studies of mammals use marks, and can therefore record repeated observations of individual's molt status (e.g. Zimova et al. 2022).
Repeat observations can provide direct observations of the speed of molt in an individual (e.g. Mumme et al. (2021b); but see also Bensch and Grahn (1993) and Rohwer and Broms (2012)), in particular when continuous molt score data is available.
This can greatly facilitate identifiability of \(\mu\) and \(\tau\).

At its core the Underhill-Zucchini model estimates the distribution of molt start dates in a population.
It is therefore conceptually straightforward to accommodate repeat measures of individuals by treating individual-level molt start dates as random intercepts that follow that very distribution.

We propose a recaptures model which allows for heterogeneity in start dates \(\mu\) but assumes a common molt duration \(\tau\).
We focus here on models for continuous molt score data (types 2-5), as even small numbers of recaptures can provide a large amount information.

Following \eqref{eq:mu-linpred} an individual's start date \(\mu_n\) then becomes

\begin{equation}
\mu_n = \mu_0 + \mu'_n + \mathbf{x}_\mu\boldsymbol{\beta}_\mu \quad\label{eq:mu-raneff}
\end{equation}

where \(\boldsymbol{x}_\mu\) is a row vector containing the values of individual-specific predictors (in the same format as \(\boldsymbol{X}_\mu\)), and \(\mu'_n\) is an individual-level random effect intercept

\begin{equation}
\mu'_n \sim \mathrm{Normal}(0,\sigma_n) \quad\label{eq:mu-raneff-dist}
\end{equation}
where \(\sigma_n\) is the individual-specific standard deviation defined in \eqref{eq:sigma-linpred}.
When repeated molt score measurements of the same individual are available during active molt, we can exploit the linearity assumption and analogous to \eqref{eq:qty} treat observed molt scores \(y_{ni}\) on dates \(u_{ni}\) as
\begin{equation}
 u_{ni} \sim \mathrm{Normal}(\mu_0 + \mu'_n + \tau y_{ni}, \sigma_\tau) \quad\label{eq:mu-raneff-obs}
\end{equation}
where \(i\) indexes the \(i\)th observation on individual \(n\), and \(\sigma_\tau\) captures any unmodelled variance in \(\tau\) as well as any measurement error in \(y\).
While recaptures within the non-molting categories hold little additional information value, recaptures across molt categories can provide additional information (Wang 2020), and the corresponding likelihoods for non-molt observations are based on the relevant cumulative distribution functions (Eqns. \eqref{eq:Pt},\eqref{eq:Rt}, Twisk and Rijmen 2009).
We refer to this model variant as the \emph{recaptures} model below, and denote it with the letter R in abbreviations, e.g.~the type 2 recaptures model (T2R).
The recaptures framework conceptually extends to the type 1 model for categorical molt data, however, exploratory simulations for this study demonstrated that the type 1 recaptures model typically requires a large number of recaptures per individual to be well identified (unpublished results).
This is in agreement with the general finding that parameter estimation in the type 1 model typically requires larger sample sizes than estimation in model types using continuous molt scores (Boersch-Supan 2021a).

\hypertarget{simulation-studies}{%
\subsection*{Simulation studies}\label{simulation-studies}}
\addcontentsline{toc}{subsection}{Simulation studies}

We use simulation studies, motivated by real-world data, to demonstrate when violations of the standard molt models lead to biased inferences, and to demonstrate that our extensions of the Underhill-Zucchini modelling framework are useful to improve inferences in these conditions.
We use relative bias compared to the simulated parameter values, and coefficient of variation (CV) of the posterior estimate as measures of accuracy and precision, respectively.
As the molt start date (day of year) is a circular variable, its absolute value and hence relative measures can be arbitrarily scaled.
We therefore calculated bias and precision for this parameter relative to the reference value for the molt duration to obtain an ecologically meaningful measure.

\hypertarget{misclassification-of-non-molting-individuals}{%
\subsubsection*{Misclassification of non-molting individuals}\label{misclassification-of-non-molting-individuals}}
\addcontentsline{toc}{subsubsection}{Misclassification of non-molting individuals}

This simulation study is motivated by a dataset presented in Erni et al. (2013), which features molt records of Southern Masked Weavers (\emph{Ploceus velatus}) from the Western Cape region of South Africa.
In this dataset birds in apparent new and apparent old plumage are recorded throughout the year, even though observations of active molt are reasonably well constrained to a six-month period.
This results in biologically implausible inferences from the type 2 molt model, leading Erni et al. (2013) to recommend the use of the type 3 model, even though this model can suffer from weak identifiability, especially when the population standard deviation of the start date is large relative to the molt duration (Underhill and Zucchini 1988; Supp. Fig. \ref{fig:supp-fig-t3-identifiability}).

We recreate this dataset by simulating molt records based on molt parameters for Southern Masked Weavers (start date 14 Jan, duration 75 days, start date std. dev. 29 days; Oschadleus (2005)), and randomly misclassifying a pre-defined proportion of non-molt records (i.e.~a subset of ``old'' records is changed to ``new'' and vice versa).
Each simulation was based on 156 randomly drawn sampling dates throughout the calendar year, with seven capture events per sampling day, yielding 1092 molt records per simulation.
Sampling of the virtual population was without replacement, so each molt record originated from a unique virtual individual.
We repeated this procedure for a total of eight misclassification rates between 0\%-20\%, and for 100 replicate simulations of the unmodified molt data.
Molt parameters were then estimated from the simulated data using the T1, T1L, T2, T2L, and T3 models, and the obtained parameter estimates and their estimated precision were compared.

\hypertarget{parameter-estimation-in-the-presence-of-molt-status-dependent-sampling-bias}{%
\subsubsection*{Parameter estimation in the presence of molt-status dependent sampling bias}\label{parameter-estimation-in-the-presence-of-molt-status-dependent-sampling-bias}}
\addcontentsline{toc}{subsubsection}{Parameter estimation in the presence of molt-status dependent sampling bias}

Not accounting for uneven sampling can lead to biased inferences about molt parameters (Bonnevie 2010a).
We here demonstrate that information in recaptures of actively molting individuals can be exploited to overcome these biases, using three simulated data sets.

\hypertarget{constant-molt-dependent-sampling-bias}{%
\paragraph*{Constant molt-dependent sampling bias}\label{constant-molt-dependent-sampling-bias}}
\addcontentsline{toc}{paragraph}{Constant molt-dependent sampling bias}

The first set of simulations assumes that molting individuals have a lower capture probability than non-molting individuals, regardless of molt progress.
We simulate a virtual population (molt start date on year day 150, molt duration 65 days, start date std. dev. 10 days) sampled on 104 days with 5 capture events per sampling day for a total of 520 molt records.
Sampling of individuals was conducted with replacement and recapture rates of individuals were allowed to vary by varying the size of the virtual super-population between 50 and 1000 individuals.
To ensure an even spread of simulation scenarios across the simulation parameter space we used a Latin square approach (Carnell 2022) to create 100 sets of simulations where capture probabilities of molting individuals ranged between 0.1-1.0 relative to non-molting individuals, and recapture rates of individuals across the population ranged between 0\%-20\%.

These parameters mean that within each resulting data set, there are on average between 2.1 and 10.4 observations per recaptured individual within a season.
Molt parameters were then estimated from the simulated data using the T1, T1R, T2, and T2R models and compared to the corresponding simulation parameters.
To visualise bias in the estimated molt parameters across the simulation parameter space we fitted generalized additive models of the general form
\emph{relative bias in molt parameter} \textasciitilde{} \emph{model type} + te(\emph{recapture rate}, \emph{relative capture probability}, by = \emph{model type})
where te() denotes a tensor product smooth with four degrees of freedom per dimension (Wood 2017).
GAM predictions that were significantly different from zero were then plotted against the simulation parameters.

\hypertarget{non-constant-molt-dependent-sampling-bias}{%
\paragraph*{Non-constant molt-dependent sampling bias}\label{non-constant-molt-dependent-sampling-bias}}
\addcontentsline{toc}{paragraph}{Non-constant molt-dependent sampling bias}

In reality, sampling biases are likely to be more complex, as the impact of molt on individuals' behavioural capabilities and/or detectabilities vary during the progress of active molt.
For example, wing raggedness in birds (i.e.~the size of the gap in the wing created by shed and incompletely regrown feathers (Bensch and Grahn 1993)) is largest at intermediate molt progress, which implies that birds with intermediate molt scores are the least able to fly and hence potentially least likely to be captured.
Such a bias towards records of birds in the early and late stages of molt is not uncommon in real-world molt datasets, and our second set of simulations emulates the sampling bias found in a citizen science dataset of molt scores from Eurasian Siskins (\emph{Spinus spinus}; Insley et al. (in prep)).
The virtual population for this set of simulations (start date year day 197, duration 78 days, start date std. dev. 9) was sampled on 25 occasions between year day 150 and year day 300 with 30 capture events per occasion.
Capture probabilities of simulated individuals were a function of their molt status (Fig. \ref{fig:siskin-capture-probabilities-figure}).
This reflects the targeted sampling approach of the emulated real-world data set.
Recapture rates were allowed to vary using the same Latin square approach as above,
but we additionally explored the sensitivity of molt models to individual heterogeneity in the molt duration \(\tau\), by allowing individual-specific molt durations \(\tau_n\) drawn from a Normal distribution with a coefficient of variation in the molt duration \(\frac{\sigma_\tau}{\tau}\) between 0\%-10\%.
We used a total of 200 simulation scenarios with recapture rates of individuals across the population between 1\%-100\% and an average of 2.0-13.4 observations per recaptured individual.
Molt parameters were then estimated from the simulated data using the T3, T3R, T5, and T5R models and compared to the corresponding simulation parameters.
Bias in the estimated molt parameters was interpolated across simulation scenarios using GAMs and visualised as described above.

\hypertarget{mixed-record-types-for-active-molt}{%
\paragraph*{Mixed record types for active molt}\label{mixed-record-types-for-active-molt}}
\addcontentsline{toc}{paragraph}{Mixed record types for active molt}

A third kind of sampling bias can arise when active molt is recorded in multiple ways. This case is motivated by the British and Irish Ringing Scheme (Baillie et al. 1999) which allows recording of active molt using the six-point feather score scale as well as the simpler categorical markers.
As a result, active molt records may consist of a mixture of categorical and score data, and excluding either data type to fit the type 1 or 2 model can affect the quantity of active molt records relative to non-molt records (as the latter can be converted from one scoring system to the other without loss of information).
The third set of simulations therefore is based on datasets simulated as molt scores of which a pre-defined proportion of active molt score records are degraded to categorical molt records.
Molt observations were simulated as for the misclassified records case, and simulated data sets were degraded by recoding between 0\%-100\% of molt scores to categorical molt codes.
Molt parameters were then estimated from the simulated data using the T1, T2, T3, and T12 models.
We also fitted the type 2 model to data rebalanced by subsampling. For this we followed the approach proposed by Bonnevie (2010a), and randomly discarded non-moult records until the proportion of non-moult to active molt records matched the proportion of days inside and outside the expected molt period across the sampling period, based on established molt durations for the relevant target species. We denote parameter estimates from this approach T2S (type 2 subsampled).
The obtained parameter estimates and their estimated precision were compared across all fitted models.

\hypertarget{real-world-data-example-american-redstart-molt}{%
\subsection*{Real-world data example: American Redstart molt}\label{real-world-data-example-american-redstart-molt}}
\addcontentsline{toc}{subsection}{Real-world data example: American Redstart molt}

Having established the performance of the extended molt models using simulations, we briefly demonstrate that these models are also suitable for real-world datasets.
For this purpose we used primary molt data from American Redstart \emph{Setophaga ruticilla} (Mumme et al. 2021a).
These data were collected during summer and fall banding operations, 1986-2000, in southwestern Pennsylvania, USA.
Details about the data collection are given in Mumme et al. (2021b), who noted that these data did not fulfil the assumptions underlying standard Underhill-Zucchini models.
The data consist of 428 sets of primary molt scores from 344 individuals, of which 42 individuals were captured at least twice within a single year.
The bulk of the molt scores are from individuals in active moult, we therefore used T3 and T3R models to analyse the data, and compared parameter estimates between the standard T3 model and the recaptures T3R model.
A worked example providing R code for this analysis is included in the Supplementary Materials.

\hypertarget{results}{%
\section*{Results}\label{results}}
\addcontentsline{toc}{section}{Results}

Bayesian estimation of molt parameters for standard molt models provided comparable results to the existing maximum likelihood method by Erni et al. (2013) (Supplementary Materials \ref{sec:std-model-appendix}). The comparative performance of standard and extended molt models under different sampling and recording scenarios is detailed below.

\hypertarget{misclassification-of-non-molting-individuals-1}{%
\subsection*{Misclassification of non-molting individuals}\label{misclassification-of-non-molting-individuals-1}}
\addcontentsline{toc}{subsection}{Misclassification of non-molting individuals}

\begin{figure}
\includegraphics[width=6.5in]{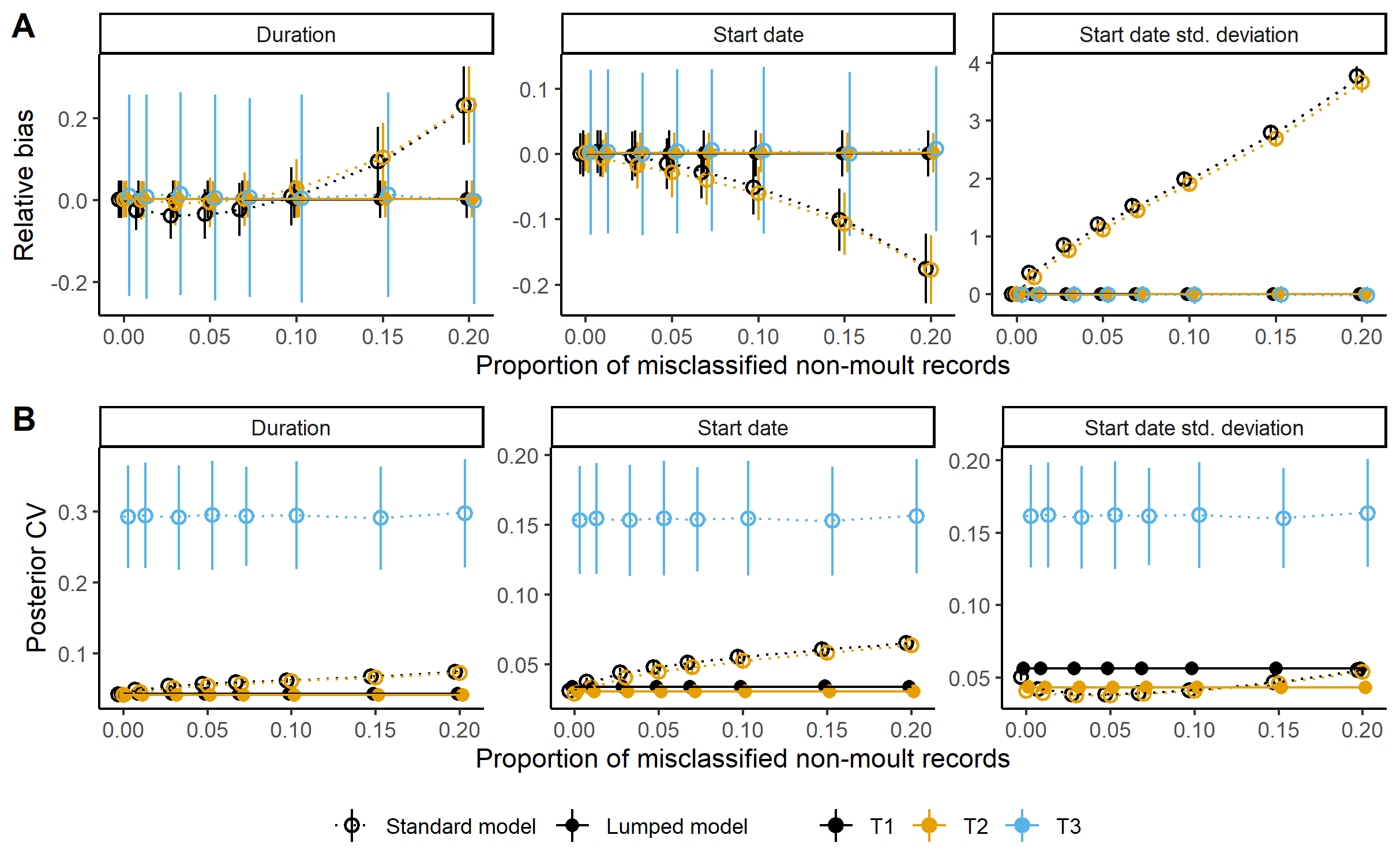} \caption{A: Even a small proportion of misclassified non-molting individuals can bias parameter estimates from the standard Type 1 and 2 models. The Type 3, and lumped Type 2 models offer unbiased estimates in the presence of misclassification. B: Among the unbiased models, the lumped Type 2 model provides more precise estimates compared to the standard Type 3 model which only takes actively molting individuals into account.  Points show mean estimates for 100 simulation runs.}\label{fig:misclass-simulations}
\end{figure}

Simulation results showed that estimates of the start date standard deviation \(\sigma\) using the T1 and T2 models were biased upwards when even a small proportion of non-molting individuals was misclassified. Estimates of duration \(\tau\) were initially less strongly affected, showing a small negative bias under model T1 and negligible bias under model T2 for misclassification rates \textless10\%. However, beyond a misclassification rate of 10\% estimates from both models were positively biased. Estimates of the start date \(\mu\) were increasingly negatively biased with increasing misclassification rates.
Estimates of all parameters obtained from both lumped model types T1L and T2L, and the T3 model (which ignores no-molt records) were unbiased (Fig. \ref{fig:misclass-simulations}A).
Lumped models (T1L,T2L) offered slightly poorer precision than their unlumped equivalents in the absence of misclassification, but their precision was otherwise unaffected by the degree of misclassification, and in all cases their precision was better than that of the corresponding T3 model (Fig. \ref{fig:misclass-simulations}B).

\hypertarget{constant-molt-dependent-sampling-bias-1}{%
\subsection*{Constant molt dependent sampling bias}\label{constant-molt-dependent-sampling-bias-1}}
\addcontentsline{toc}{subsection}{Constant molt dependent sampling bias}

\begin{figure}
\includegraphics[width=7in]{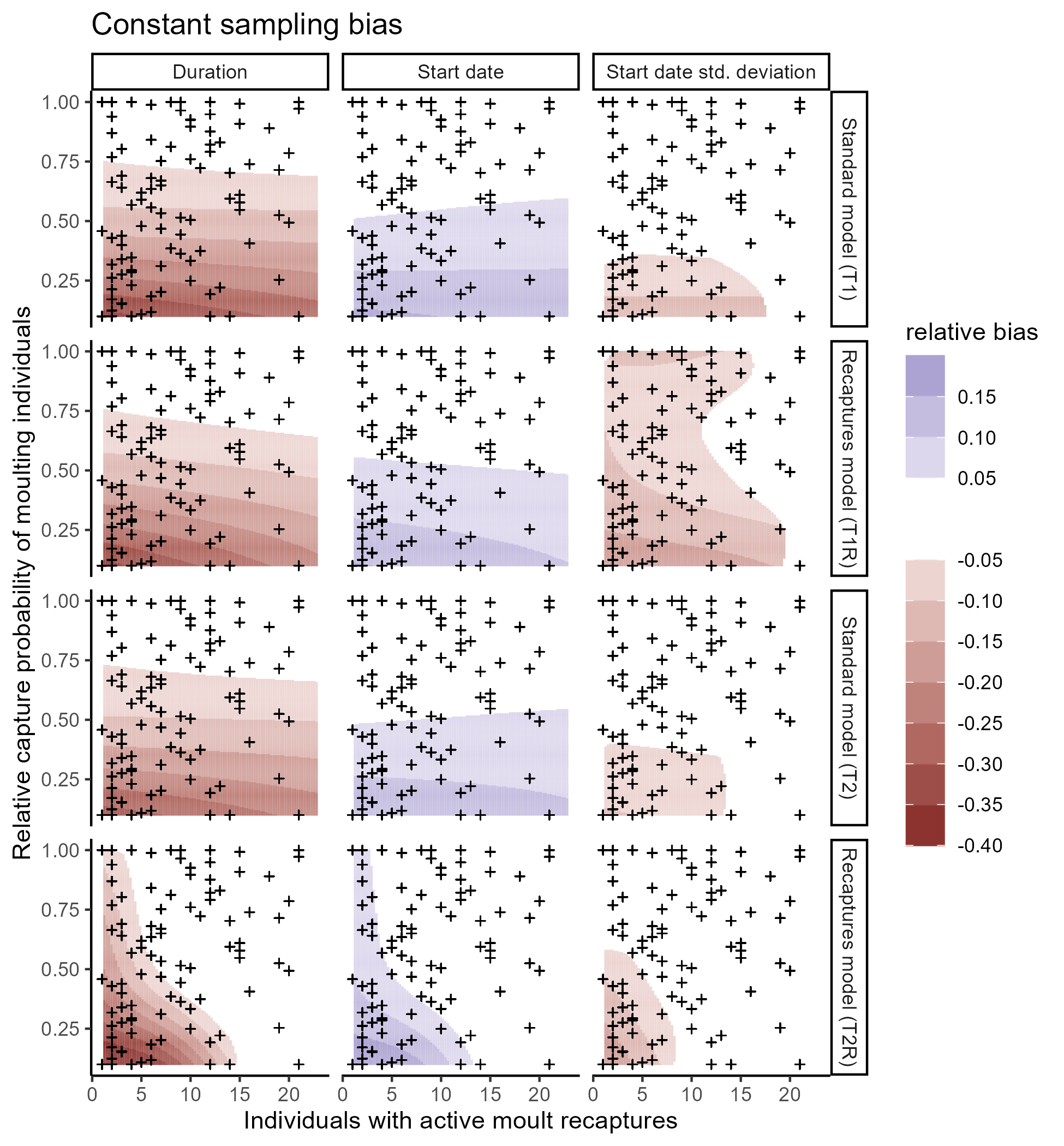} \caption{When capture probabilities of actively molting individuals are lower than those of non-molting individuals, inferences from the standard Underhill-Zucchini models (rows 1,3) may be biased, and the bias increases as relative capture probabilities of molting individuals decrease. Recapture information can help overcome this bias, although low numbers of recaptures per individual only provide negligible gains for the T1R model (row 2). However, for the T2R model (row 4) bias decreases rapidly as the number of individuals with repeat observations during active molt increases. Bias estimates were interpolated using a GAM, crossmarks indicate the positions of 100 simulated datasets in parameter space.}\label{fig:constant-sampling-bias-sims}
\end{figure}

Simulation results showed that estimates of all parameters from the T1 and T2 models were affected when the sampling probability of molting individuals dropped below 50\%-75\% relative to non-molting individuals (Fig. \ref{fig:constant-sampling-bias-sims}).
The type 2 recapture model (T2R) was able to overcome this bias, but the stronger the sampling bias, the more individuals with recaptures during active molt were needed to obtain unbiased parameter estimates.
Parameter estimates were generally unbiased whenever active-molt recaptures from 10 or more individuals were available.
Gains from using the type 1 recaptures model (T1R) were negligible for recapture rates comparable to typical mark-recapture datasets obtained from physical capture methods, although the T1R model outperformed the standard T1 model at high individual recapture rates (Supplementary Materials \ref{sec:t1r-appendix}).

\hypertarget{non-constant-molt-dependent-sampling-bias-1}{%
\subsection*{Non-constant molt dependent sampling bias}\label{non-constant-molt-dependent-sampling-bias-1}}
\addcontentsline{toc}{subsection}{Non-constant molt dependent sampling bias}

\begin{figure}
\includegraphics[width=7in]{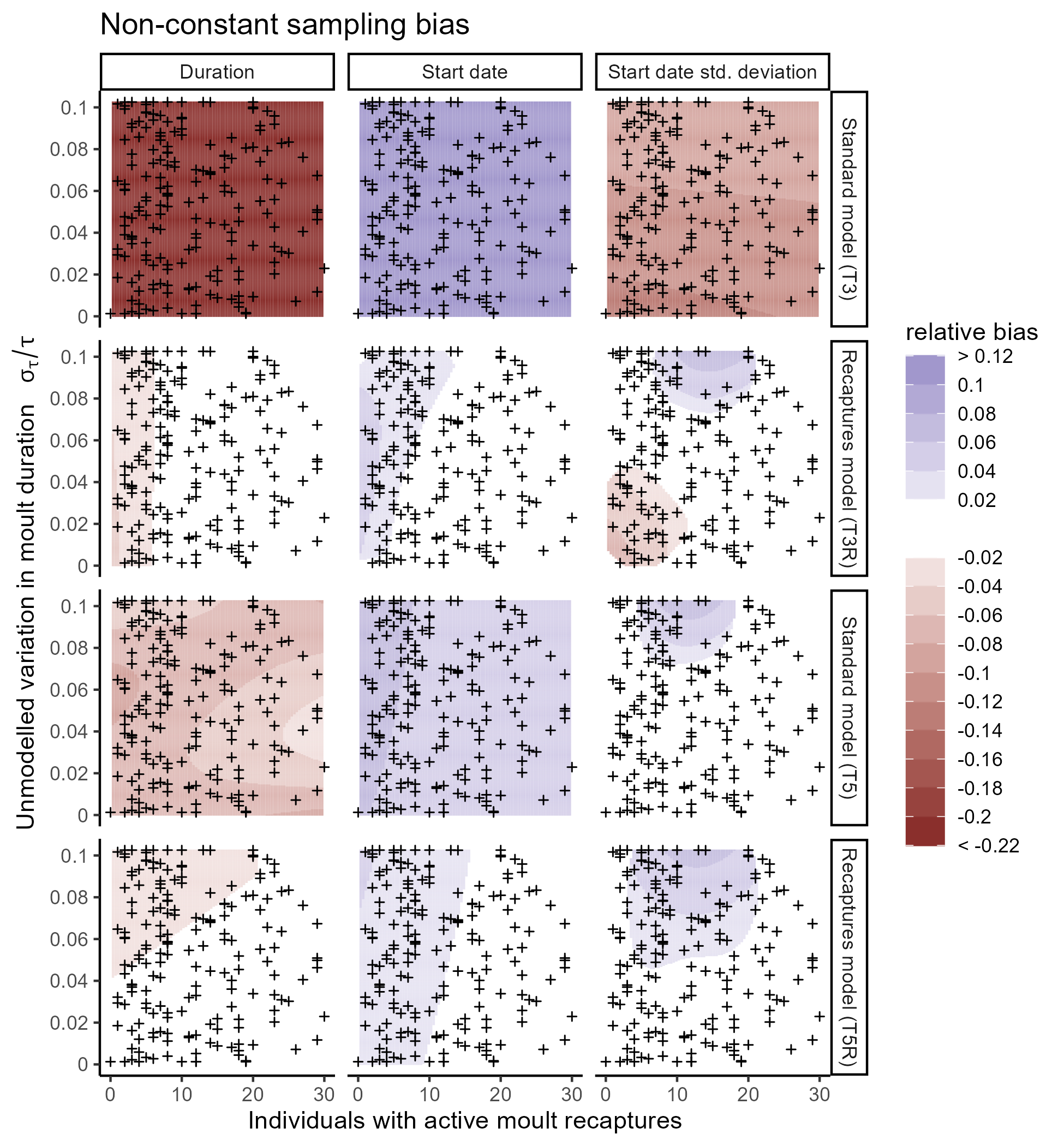} \caption{When sampling of actively molting individuals is biased towards individuals in early molt, inferences from the standard Underhill-Zucchini type 3 and type 5 model (1st and 3rd row) are biased, and the bias increases in the presence of unmodelled variation $\sigma_\tau$ in the molt duration parameter $\tau$. The corresponding recaptures models (2nd and 4th row) are much less biased, and bias decreases rapidly as the proportion of individuals with repeat observations during active molt increases. Bias estimates were interpolated using a GAM, crossmarks indicate the positions of 100 simulated datasets in parameter space.}\label{fig:siskin-t3-simulations}
\end{figure}

Using simulated data with a similar sampling bias to a real-world dataset we further demonstrate that inferences from the recaptures model were much less biased than inferences from the standard model, and that bias decreased with an increasing proportion of individuals with repeated observations during active molt (Fig. \ref{fig:siskin-t3-simulations}).
Parameter estimates for were generally unbiased whenever active-molt recaptures from 5 or more individuals were available for the T3R model, and 10 or more individuals for the T5R model.
We also show that unmodelled individual heterogeneity in the molt duration \(\tau\) exacerbated biases and that a the proportion of individuals with recaptures required to obtain unbiased parameter estimates approximately doubled when the unmodelled variation in the duration increased from 0\% to 10\%.

\hypertarget{mixed-record-types-for-active-molt-1}{%
\subsection*{Mixed record types for active molt}\label{mixed-record-types-for-active-molt-1}}
\addcontentsline{toc}{subsection}{Mixed record types for active molt}

\begin{figure}
\includegraphics[width=6.5in]{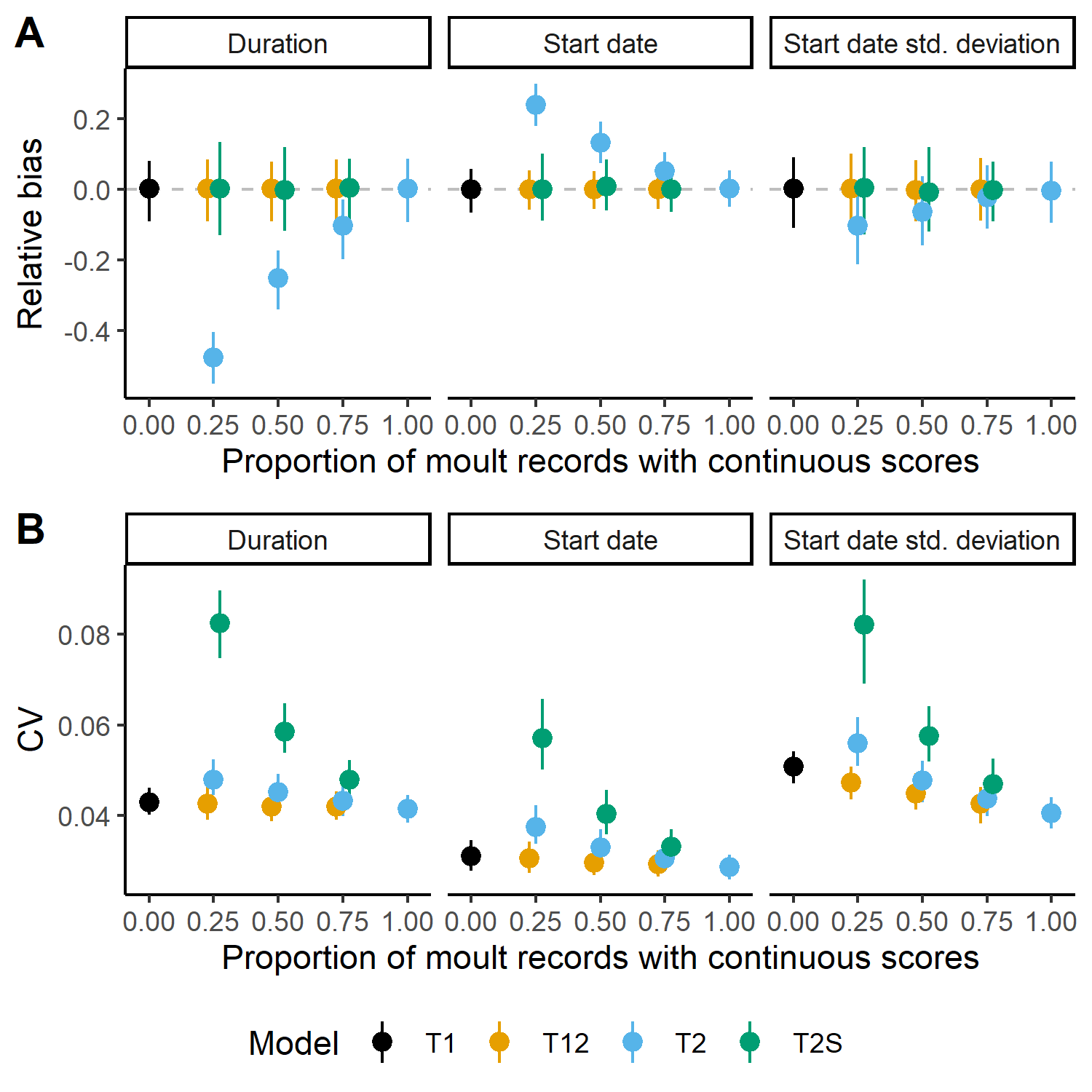} \caption{Integrating categorical and feather score data in the same model allows the analysis of data sources with different recording protocols. A: Type 2 models omitting categorical records of molting individuals but retaining all non-molt records (T2) yield increasingly biased parameter estimates as the proportion of omitted data increases. Both the integrated type 12 model (T12), and the type 2 model fitted to data rebalanced by subsampling (T2S) yield unbiased estimates. B: The type 12 model delivers a similar level of precision as the type 1 and type 2 models fitted to the full dataset. Subsampling the data to restore balance among categories (T2S) degrades the precision of parameter estimates.}\label{fig:euring-integration}
\end{figure}

Discarding molt records with incomplete active molt scores can lead to similar parameter biases in the T2 model as biased sampling of molting individuals.
Simulation results showed that the re-balancing through subsampling allows unbiased estimates from the T2 model (Fig. \ref{fig:euring-integration}A), but the precision of those estimates suffers because subsampling reduces the sample size available to the model(Fig. \ref{fig:euring-integration}B).
Using all available data by including categorical molt records allows unbiased inferences with a precision that is at least as good, if not better than the corresponding T1 model (Fig. \ref{fig:euring-integration}).

\hypertarget{real-world-data-example-american-redstart-molt-1}{%
\subsection*{Real-world data example: American Redstart molt}\label{real-world-data-example-american-redstart-molt-1}}
\addcontentsline{toc}{subsection}{Real-world data example: American Redstart molt}

\begin{figure}
\includegraphics[width=6.5in]{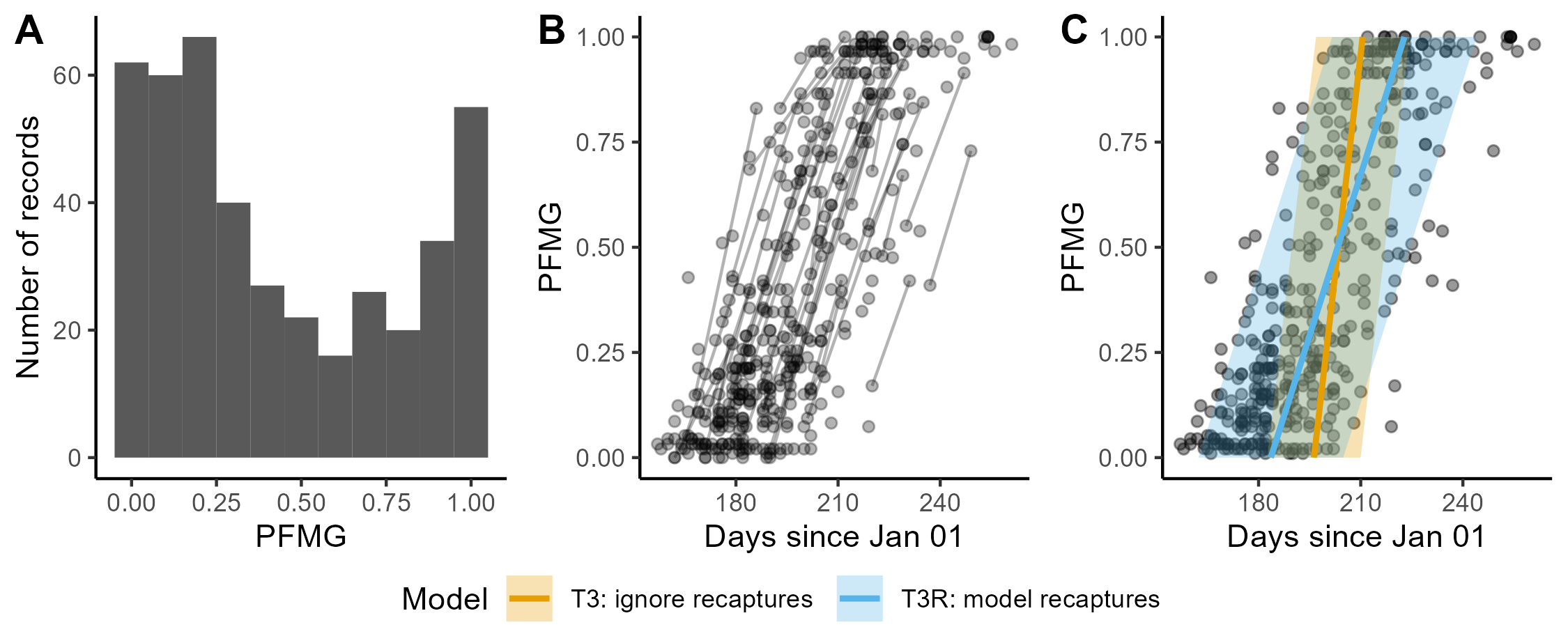} \caption{Extended molt phenology models improve inferences about molt duration and timings from real-world molt data. A: Molt records from American Redstart observations are biased towards individuals in early or late molt. B: Seasonal progression of observed American Redstart molt scores. Lines connect observations from recaptured individuals. C: The T3 molt model converge at biased parameter estimates (orange lines and polygons). When including recapture information (T3R model), the resulting fit (blue lines and polygons) better encapsulates the observed data. PFMG: proportion of feather mass grown.}\label{fig:amre-example}
\end{figure}

As noted by Mumme et al. (2021b), the American Redstart data do not fulfill the assumption of homogeneous capture probabilities, as individuals at intermediate molt progress are underrepresented in the data (Fig. \ref{fig:amre-example}A).
As a consequence a naive T3 fit to the data yields biologically non-sensical parameter estimates, with an estimated molt duration of 14.3 days (95\%CI 10.8-20.4), and a molt start date on July 15 (year day 196; 95\%CI 193-199), well after most observed birds have commenced moult, and a population standard deviation of the start date of 6.99 days (6.30-8.22; Fig. \ref{fig:amre-example}B,C).
As a consequence, the estimated 95\% molt interval (i.e.~the time interval in which 95\% of the population are expected to be in active moult, defined as the polygon bounded by the mean molt trajectory \(\pm\) 1.96 \(\times\) Start date SD; orange shaded area in Fig. \ref{fig:amre-example}C) only encapsulates a fraction of the observed molt records.
In contrast, the T3R recaptures model estimates a molt duration of 39.3 days (37.3-41.1), a mean start date on July 3 (year day 184; 95\%CI 183-185), and a population standard deviation of the start date of 10.8 days (9.98-11.8; Fig. \ref{fig:amre-example}B,C).
The corresponding 95\% molt interval (blue shaded area in Fig.\textasciitilde\ref{fig:amre-example}C) encapsulates the majority of observations, suggesting a much more appropriate model fit.
The T3R estimates also show good agreement with the estimates for molt duration and midpoint of primary molt reported by Mumme et al. (2021b) (Tab.\textasciitilde\ref{tab:estimates-table}).

\begin{table}

\caption{\label{tab:estimates-table}Moult parameter estimates for American Redstart.}
\centering
\begin{tabular}[t]{llll}
\toprule
Parameter & This study (T3) & This study (T3R) & Mumme et al. (2021)\\
\midrule
Molt duration (days) & 14.3 (2.58) & 39.3 (0.97) & 41 (N/A)\\
Start date & July 15 (1.7 days) & July 3 (0.6 days) & N/A\\
Midpoint of primary molt & July 22 (0.9 days) & July 23 (0.2 days) & July 19 (0.7 days)\\
Start date SD (days) & 6.99 (0.5) & 10.8 (0.5) & N/A\\
\bottomrule
\end{tabular}
\end{table}

\clearpage

\hypertarget{discussion}{%
\section*{Discussion}\label{discussion}}
\addcontentsline{toc}{section}{Discussion}

Understanding the mechanisms underlying the timing and duration of molt and its position in the annual cycle is crucial to understanding birds' and mammals' responses to a changing environment (Watson 1963, Hällfors et al. 2020, Hanmer et al. 2022).
Our work contributes to this goal by expanding existing molt models, in particular by adding the option of exploiting information contained in mark-recapture data.
This allows a relaxation of assumptions about random sampling of molting individuals which are inherent in existing molt models, but are often difficult to meet under field conditions.
We demonstrate that violations of the assumptions of standard molt models can lead to biased inferences about molt phenology.
The extended models we present here are more robust to non-random sampling of individuals, and thus allow better inferences about molt phenology from imperfect, real-world datasets.

Our extended molt models are particularly useful when repeated sampling of individuals is possible, which has important consequences for the design of molt phenology studies.
As avian molt observations are often collected in conjunction with ringing, the identification of recaptured individuals is in principle possible, and although within-season recaptures are generally rare in avian molt datasets (Ginn and Melville 1983, Rose et al. 2020, Mumme et al. 2021b), we demonstrate both through simulation and with a real-world dataset that even a small number of recaptured individuals can drastically improve the accuracy and precision of molt parameter estimates.
In the case of the real-world American Redstart example, our molt parameter estimates from the extended Underhill-Zucchini are similar to the estimates derived by Mumme et al. (2021b) using information from the recaptures alone for the molt duration, and an ad-hoc approach to estimate the molt start date (Tab. \ref{tab:estimates-table}).
However, unlike the ad-hoc approach, our extended Underhill-Zucchini model estimates the molt start date, its population standard deviation, and the molt duration simultaneously and therefore consistently estimates the correlated uncertainties in these three parameters.
Robustly estimating these three parameters, is important when trying to understand population-level changes in the phenology of molt, as changes in the full phenological distribution of molt at the population-level can arise from both individual-level and population level effects which may have different mechanistic drivers and/or consequences for intra- and interspecific ecological interactions (Hällfors et al. 2020, Hanmer et al. 2022, Macphie et al. 2023).
Additional applications of both the lumped and recaptures molt models to real-world data are presented in Boersch-Supan et al. (2023b), who demonstrate that these models can address uncertainty about the assignment of non-moulting birds, and that even small proportions of recaptures (\textasciitilde1\% of c.~4300 captured individuals) can sufficiently constrain molt models to improve parameter precision and mitigate against parameter biases caused by uneven sampling.
Further empirical research is required to better understand how much recapture data is required to overcome molt-status dependent sampling biases.
A key requirement for robust inferences about molt durations and start dates using the recaptures models is to estimate both the duration parameter and the population standard deviation of the start date well.
Our simulations demonstrated that this can be the case with within-molt recaptures from as few as 5-10 individuals, however, larger sample sizes are required when unmodelled individual heterogeneity in the molt duration is present.
Additionally, while our modelling approach addresses heterogeneity in capture probabilities with respect to molt status, it relies - as many mark-recapture methods - on the assumption that (re)captured individuals are representative of the sampled population as a whole (see e.g. Nichols et al. 1984, Cubaynes et al. 2010, Abadi et al. 2013).
This assumption, i.e.~that all individuals of a certain molt status are equally catchable, may not always hold, and it is therefore important to assess whether the sample of recaptured individuals differs in any potentially relevant characteristics from the sample of not recaptured individuals.

For practitioners collecting and analysing real world molt data sets we strongly suggest that (i) individual feather scores of all encountered molting and non-molting birds are recorded, (ii) that encountered birds are marked and that a sampling design that aims to maximise the probability of recapturing a proportion of marked individuals within the active molt period is used, and (iii) that any fitted model is visually checked, e.g.~by plotting the 95\% molt interval against the observations using the \textbf{moult\_plot} function in the \textbf{moultmcmc} package.
The molt interval represents the time interval in which 95\% of the population are expected to be in active molt, and is defined as the polygon bounded by the mean molt trajectory \(\pm\) 1.96 \(\times\) Start date SD (Fig. \ref{fig:amre-example}C).
This interval should encapsulates a large proportion of the observed molt records (albeit not necessarily 95\% of the observations, as this interval is not strictly equivalent to the prediction interval of a linear regression model).
For more formal comparisons, expected daily frequencies of molt observations under the model can be computed and compared with the observed frequencies following the approach outlined in Underhill and Zucchini (1988).

Less invasive re-encounter methods, such as camera trapping, may be limited to providing less information-rich categorical molt data, but at the same time such methods can provide sufficiently high recapture rates to identify individual-level molt parameters where individuals are identifiable.
Such data are currently more common in mammalian systems (e.g. Zimova et al. 2020a), but may become increasingly available for birds (O'Brien and Kinnaird 2008, Pyle 2022).

Although our study focusses on molt, the modelling framework we present is applicable to a wide range of ontogenetic and/or seasonal life-cycle transitions across many animal and plant taxa, as long as they can be recorded using either ordered categories or approximately linear progress scores between well-defined start and end states (e.g. Gosner 1960, Lancashire et al. 1991, Redfern 2010).
As a consequence of this linearity assumption our models therefore do not currently address more complex life-cycle transitions, such as - in the avian molt context - arrested moult, suspended moult, partial moult or Staffelmauser.
However, our modelling framework is in principle further extendable to address these more complex phenomena, e.g.~analogous to phenology models applied to invertebrate life-cycles (see e.g. Candy 1991), albeit at the expense of estimating additional parameters.
Our implementation of the recaptures models further assume that all heterogeneity in molt duration is captured by relevant covariates, i.e.~we do not allow for the separate estimation of the population variation in the molt duration and individual variation in the molt duration, and/or measurement errors in the molt scores of recaptured individuals - these quantities are confounded in the \(\sigma_{\tau}\) parameter.
Models to separately estimate these quantities can be derived in principle (e.g. Wang 2020), but robust parameter estimation, particularly in the presence of measurement error, requires much higher numbers of within-season recaptures per individual (ideally \(\gg3\)) than we believe are achievable for most molt studies in wild bird populations, given that it is excedingly rare that more than two within-season recaptures per individual are recorded in real-world molt data sets (e.g. Ginn and Melville 1983, Rose et al. 2020, Mumme et al. 2021b, Boersch-Supan et al. 2023b).
Improved inferences about molt and other life-cycle events do not only allow for a better understanding of phenological cycles per se, but also processes linked to them, such as disease dynamics (Benedict et al. 2023), as well as informing the timing of conservation and management actions (Brom et al. 2023).
We hope that our model extensions and their implementation in the R package \emph{moultmcmc} (Boersch-Supan et al. 2023a) will continue to lower the hurdle of applying phenological models to empirical data describing molt and other life-cycle events.

\hypertarget{data-availability}{%
\section*{Data availability}\label{data-availability}}
\addcontentsline{toc}{section}{Data availability}

Data and code for the simulation and case studies are archived on Zenodo: \url{https://doi.org/10.5281/zenodo.7643922}

\hypertarget{references}{%
\section*{References}\label{references}}
\addcontentsline{toc}{section}{References}

\hypertarget{refs}{}
\begin{CSLReferences}{1}{0}
\leavevmode\vadjust pre{\hypertarget{ref-abadi2013revisiting}{}}%
Abadi, F., A. Botha, and R. Altwegg (2013). \href{https://doi.org/10.1371/journal.pone.0062636}{Revisiting the effect of capture heterogeneity on survival estimates in capture-mark-recapture studies: Does it matter?} PLOS ONE 8:1--8.

\leavevmode\vadjust pre{\hypertarget{ref-baillie1999development}{}}%
Baillie, S. R., C. V. Wernham, and J. A. Clark (1999). Development of the {British} and {Irish} ringing scheme and its role in conservation biology. Ringing \& Migration 19:5--19.

\leavevmode\vadjust pre{\hypertarget{ref-beltran2018convergence}{}}%
Beltran, R. S., J. M. Burns, and G. A. Breed (2018). Convergence of biannual moulting strategies across birds and mammals. Proceedings of the Royal Society B: Biological Sciences 285:20180318.

\leavevmode\vadjust pre{\hypertarget{ref-beltran2019reproductive}{}}%
Beltran, R. S., A. L. Kirkham, G. A. Breed, J. W. Testa, and J. M. Burns (2019). Reproductive success delays moult phenology in a polar mammal. Scientific Reports 9:1--12.

\leavevmode\vadjust pre{\hypertarget{ref-benedict2023sores}{}}%
Benedict, B. M., P. S. Barboza, J. A. Crouse, K. R. Groch, M. R. Kulpa, D. P. Thompson, G. G. Verocai, and D. J. Wiener (2023). \href{https://doi.org/10.1371/journal.pone.0278886}{Sores of boreal moose reveal a previously unknown genetic lineage of parasitic nematode within the genus \emph{{O}nchocerca}}. PLoS ONE 18:1--14.

\leavevmode\vadjust pre{\hypertarget{ref-bensch1993new}{}}%
Bensch, S., and M. Grahn (1993). A new method for estimating individual speed of molt. The Condor 95:305--315.

\leavevmode\vadjust pre{\hypertarget{ref-boerschsupan2021feather}{}}%
Boersch-Supan, P. (2021a). A feather at a time: Moult recording. LifeCycle 10:36--37.

\leavevmode\vadjust pre{\hypertarget{ref-boerschsupan2021modeling}{}}%
Boersch-Supan, P. H. (2021b). \href{https://doi.org/10.5281/zenodo.5006005}{{{[}Re{]} Modeling Insect Phenology Using Ordinal Regression and Continuation Ratio Models}}. ReScience C 7.

\leavevmode\vadjust pre{\hypertarget{ref-moultmcmc}{}}%
Boersch-Supan, P. H., H. J. Hanmer, and R. A. Robinson (2023a). Moultmcmc v0.1.0. DOI: 10.5281/zenodo.7643725. {[}Online.{]} Available at \url{https://doi.org/10.5281/zenodo.7643725}.

\leavevmode\vadjust pre{\hypertarget{ref-boerschsupan2022demo}{}}%
Boersch-Supan, P. H., A. T. K. Lee, and H.-D. Oschadleus (2023b). \href{https://doi.org/10.2989/00306525.2023.2248396}{A demonstration of the value of recapture data for informing moult phenology models for species with imperfect moult data}. Ostrich 94:2248396.

\leavevmode\vadjust pre{\hypertarget{ref-bonnevie2010relative}{}}%
Bonnevie, B. T. (2010b). \href{https://doi.org/10.2989/00306525.2010.455820}{Relative feather mass indices: Are feather masses needed to estimate the percentage of new feather mass grown for moult regression models?} Ostrich 81:59--62.

\leavevmode\vadjust pre{\hypertarget{ref-bonnevie2010balancing}{}}%
Bonnevie, B. T. (2010a). Balancing moult data by subsampling non-moulting birds prior to regression analysis. Ostrich 81:265--268.

\leavevmode\vadjust pre{\hypertarget{ref-bridge2011mind}{}}%
Bridge, E. S. (2011). Mind the gaps: What's missing in our understanding of feather molt. The Condor 113:1--4.

\leavevmode\vadjust pre{\hypertarget{ref-brom2023mowing}{}}%
Brom, P. D., L. G. Underhill, K. Winter, and J. F. Colville (2023). A mowing strategy for urban parks to support spring flowers in a mediterranean climate city in south africa. Urban Ecosystems:1--11.

\leavevmode\vadjust pre{\hypertarget{ref-burkner2019ordinal}{}}%
Bürkner, P.-C., and M. Vuorre (2019). \href{https://doi.org/10.1177/2515245918823199}{Ordinal regression models in psychology: A tutorial}. Advances in Methods and Practices in Psychological Science 2:77--101.

\leavevmode\vadjust pre{\hypertarget{ref-candy1991modeling}{}}%
Candy, S. G. (1991). Modeling insect phenology using ordinal regression and continuation ratio models. Environmental Entomology 20:190--195.

\leavevmode\vadjust pre{\hypertarget{ref-lhs}{}}%
Carnell, R. (2022). Lhs: Latin hypercube samples. {[}Online.{]} Available at \url{https://CRAN.R-project.org/package=lhs}.

\leavevmode\vadjust pre{\hypertarget{ref-caro2005adaptive}{}}%
Caro, T. (2005). The adaptive significance of coloration in mammals. BioScience 55:125--136.

\leavevmode\vadjust pre{\hypertarget{ref-carpenter2017stan}{}}%
Carpenter, B., A. Gelman, M. D. Hoffman, D. Lee, B. Goodrich, M. Betancourt, M. Brubaker, J. Guo, P. Li, and A. Riddell (2017). \href{https://doi.org/10.18637/jss.v076.i01}{Stan: A probabilistic programming language}. Journal of Statistical Software 76:1--32.

\leavevmode\vadjust pre{\hypertarget{ref-chuine2017process}{}}%
Chuine, I., and J. Régnière (2017). Process-based models of phenology for plants and animals. Annual Review of Ecology, Evolution, and Systematics 48:159--182.

\leavevmode\vadjust pre{\hypertarget{ref-craig2014primary}{}}%
Craig, A. J., B. T. Bonnevie, P. E. Hulley, and G. D. Underhill (2014). \href{https://doi.org/10.2989/00306525.2014.931310}{Primary wing-moult and site fidelity in {South African} mousebirds ({Coliidae})}. Ostrich 85:171--175.

\leavevmode\vadjust pre{\hypertarget{ref-cubaynes2010importance}{}}%
Cubaynes, S., R. Pradel, R. Choquet, C. Duchamp, J.-M. Gaillard, J.-D. Lebreton, E. Marboutin, C. Miquel, A.-M. Reboulet, C. Poillot, et al. (2010). Importance of accounting for detection heterogeneity when estimating abundance: The case of french wolves. Conservation Biology 24:621--626.

\leavevmode\vadjust pre{\hypertarget{ref-dawson2014mammals}{}}%
Dawson, T. J., K. N. Webster, and S. K. Maloney (2014). The fur of mammals in exposed environments; do crypsis and thermal needs necessarily conflict? The polar bear and marsupial koala compared. Journal of Comparative Physiology B 184:273--284.

\leavevmode\vadjust pre{\hypertarget{ref-dennis1986stochastic}{}}%
Dennis, B., W. P. Kemp, and R. C. Beckwith (1986). \href{https://doi.org/10.1093/ee/15.3.540}{Stochastic model of insect phenology: Estimation and testing}. Environmental Entomology 15:540--546.

\leavevmode\vadjust pre{\hypertarget{ref-dolnik1974semiquantitative}{}}%
Dolnik, V., and V. Gavrilov (1974). Polukolichestvennyy metod registratsii lin'ki u vorob'inykh ptits (semiquantitative method of the molt registration in passerine birds). Ornitologia 11:110--125.

\leavevmode\vadjust pre{\hypertarget{ref-erni2013moult}{}}%
Erni, B., B. T. Bonnevie, H.-D. Oschadleus, R. Altwegg, and L. G. Underhill (2013). \href{https://doi.org/10.18637/jss.v052.i08}{Moult: An {R} package to analyze moult in birds}. Journal of Statistical Software 52.

\leavevmode\vadjust pre{\hypertarget{ref-ginn1983moult}{}}%
Ginn, H., and D. Melville (1983). Moult in birds. British Trust for Ornithology, Tring, UK.

\leavevmode\vadjust pre{\hypertarget{ref-gosner1960simplified}{}}%
Gosner, K. L. (1960). \href{http://www.jstor.org/stable/3890061}{A simplified table for staging anuran embryos and larvae with notes on identification}. Herpetologica 16:183--190.

\leavevmode\vadjust pre{\hypertarget{ref-hallfors2020shifts}{}}%
Hällfors, M. H., L. H. Antão, M. Itter, A. Lehikoinen, T. Lindholm, T. Roslin, and M. Saastamoinen (2020). Shifts in timing and duration of breeding for 73 boreal bird species over four decades. Proceedings of the National Academy of Sciences 117:18557--18565.

\leavevmode\vadjust pre{\hypertarget{ref-hanmer2022differential}{}}%
Hanmer, H. J., P. H. Boersch-Supan, and R. A. Robinson (2022). Differential changes in lifecycle-event phenology provide a window into regional population declines. Biology Letters 18:20220186.

\leavevmode\vadjust pre{\hypertarget{ref-haukioja1971flightlessness}{}}%
Haukioja, E. (1971). Flightlessness in some moulting passerines in {Northern Europe}. Ornis Fennica 48:1--1.

\leavevmode\vadjust pre{\hypertarget{ref-insley2022moult}{}}%
Insley, H., S. Beck, and P. Boersch-Supan (in prep). {Breeding and moult phenology of siskins in the Scottish Highlands}.

\leavevmode\vadjust pre{\hypertarget{ref-jenni2020biology}{}}%
Jenni, L., and R. Winkler (2020). The biology of moult in birds. Bloomsbury Publishing.

\leavevmode\vadjust pre{\hypertarget{ref-dekock2021determinants}{}}%
Kock, L. de, W. C. Oosthuizen, R. S. Beltran, M. N. Bester, and P. de Bruyn (2021). Determinants of moult haulout phenology and duration in southern elephant seals. Scientific Reports 11:1--13.

\leavevmode\vadjust pre{\hypertarget{ref-lancashire1991uniform}{}}%
Lancashire, P. D., H. Bleiholder, T. V. D. Boom, P. Langelüddeke, R. Stauss, E. Weber, and A. Witzenberger (1991). A uniform decimal code for growth stages of crops and weeds. Annals of Applied Biology 119:561--601.

\leavevmode\vadjust pre{\hypertarget{ref-ling1970pelage}{}}%
Ling, J. K. (1970). Pelage and molting in wild mammals with special reference to aquatic forms. The Quarterly Review of Biology 45:16--54.

\leavevmode\vadjust pre{\hypertarget{ref-ling1972adaptive}{}}%
Ling, J. K. (1972). Adaptive functions of vertebrate molting cycles. American Zoologist 12:77--93.

\leavevmode\vadjust pre{\hypertarget{ref-macphie2023modelling}{}}%
Macphie, K. H., J. M. Samplonius, J. L. Pick, J. D. Hadfield, and A. B. Phillimore (2023). \href{https://doi.org/10.1111/1365-2435.14436}{Modelling thermal sensitivity in the full phenological distribution: A new approach applied to the spring arboreal caterpillar peak}. Functional Ecology 00:1--12.

\leavevmode\vadjust pre{\hypertarget{ref-marra2015call}{}}%
Marra, P. P., E. B. Cohen, S. R. Loss, J. E. Rutter, and C. M. Tonra (2015). A call for full annual cycle research in animal ecology. Biology Letters 11:20150552.

\leavevmode\vadjust pre{\hypertarget{ref-matloff2020flight}{}}%
Matloff, L. Y., E. Chang, T. J. Feo, L. Jeffries, A. K. Stowers, C. Thomson, and D. Lentink (2020). How flight feathers stick together to form a continuous morphing wing. Science 367:293--297.

\leavevmode\vadjust pre{\hypertarget{ref-mcqueen2019evolutionary}{}}%
McQueen, A., B. Kempenaers, J. Dale, M. Valcu, Z. T. Emery, C. J. Dey, A. Peters, and K. Delhey (2019). Evolutionary drivers of seasonal plumage colours: Colour change by moult correlates with sexual selection, predation risk and seasonality across passerines. Ecology Letters 22:1838--1849.

\leavevmode\vadjust pre{\hypertarget{ref-moyes2011advancing}{}}%
Moyes, K., D. H. Nussey, M. N. Clements, F. E. Guinness, A. Morris, S. Morris, J. M. Pemberton, L. E. Kruuk, and T. Clutton-Brock (2011). Advancing breeding phenology in response to environmental change in a wild red deer population. Global Change Biology 17:2455--2469.

\leavevmode\vadjust pre{\hypertarget{ref-mumme2021data}{}}%
Mumme, R. L., R. S. Mulvihill, and D. Norman (2021a). {High-intensity flight feather molt and comparative molt ecology of warblers of eastern North America {[}Dataset{]}}. Dryad. https://doi.org/\url{https://doi.org/10.5061/dryad.63xsj3v0x}

\leavevmode\vadjust pre{\hypertarget{ref-mumme2021high}{}}%
Mumme, R. L., R. S. Mulvihill, and D. Norman (2021b). \href{https://doi.org/10.1093/ornithology/ukaa072}{{High-intensity flight feather molt and comparative molt ecology of warblers of eastern North America}}. Ornithology 138.

\leavevmode\vadjust pre{\hypertarget{ref-newton1966moult}{}}%
Newton, I. (1966). The moult of the bullfinch \emph{{Pyrrhula} pyrrhula}. Ibis 108:41--67.

\leavevmode\vadjust pre{\hypertarget{ref-newton2009moult}{}}%
Newton, I. (2009). Moult and plumage. Ringing \& Migration 24:220--226.

\leavevmode\vadjust pre{\hypertarget{ref-nichols1984effects}{}}%
Nichols, J. D., J. E. Hines, and K. H. Pollock (1984). \href{http://www.jstor.org/stable/3808491}{Effects of permanent trap response in capture probability on jolly-seber capture-recapture model estimates}. The Journal of Wildlife Management 48:289--294.

\leavevmode\vadjust pre{\hypertarget{ref-obrien2008picture}{}}%
O'Brien, T. G., and M. F. Kinnaird (2008). \href{https://doi.org/10.1017/S0959270908000348}{A picture is worth a thousand words: The application of camera trapping to the study of birds}. Bird Conservation International 18:S144--S162.

\leavevmode\vadjust pre{\hypertarget{ref-ogle2020ensuring}{}}%
Ogle, K., and J. J. Barber (2020). Ensuring identifiability in hierarchical mixed effects {Bayesian} models. Ecological Applications 30:e02159.

\leavevmode\vadjust pre{\hypertarget{ref-oschadleus2005patterns}{}}%
Oschadleus, H.-D. (2005). Patterns of primary moult in the weavers, {Ploceidae}. University of Cape Town.

\leavevmode\vadjust pre{\hypertarget{ref-pyle2022examination}{}}%
Pyle, P. (2022). Examination of macaulay library images to determine avian molt strategies: A case study on hummingbirds. The Wilson Journal of Ornithology 134:52--65.

\leavevmode\vadjust pre{\hypertarget{ref-redfern2010brood}{}}%
Redfern, C. P. (2010). Brood-patch development and female body mass in passerines. Ringing \& Migration 25:33--41.

\leavevmode\vadjust pre{\hypertarget{ref-rohwer2012use}{}}%
Rohwer, S., and K. Broms (2012). \href{https://doi.org/10.1525/auk.2012.12096}{{Use of Feather Loss Intervals to Estimate Molt Duration and to Sample Feather Vein at Equal Time Intervals Through the Primary Replacement}}. The Auk 129:653--659.

\leavevmode\vadjust pre{\hypertarget{ref-rose2020summarising}{}}%
Rose, S., R. L. Thomson, H.-D. Oschadleus, and A. T. Lee (2020). Summarising biometrics from the SAFRING database for southern african birds. Ostrich 91:169--173.

\leavevmode\vadjust pre{\hypertarget{ref-rothery2002simple}{}}%
Rothery, P., and I. Newton (2002). A simple method for estimating timing and duration of avian primary moult using field data. Ibis 144:526--528.

\leavevmode\vadjust pre{\hypertarget{ref-shutt2019environmental}{}}%
Shutt, J. D., I. B. Cabello, K. Keogan, D. I. Leech, J. M. Samplonius, L. Whittle, M. D. Burgess, and A. B. Phillimore (2019). The environmental predictors of spatio-temporal variation in the breeding phenology of a passerine bird. Proceedings of The Royal Society B: Biological Sciences 286:20190952.

\leavevmode\vadjust pre{\hypertarget{ref-rstan}{}}%
Stan Development Team (2021). {RStan}: The {R} interface to {Stan}. {[}Online.{]} Available at \url{https://mc-stan.org/}.

\leavevmode\vadjust pre{\hypertarget{ref-tomotani2018climate}{}}%
Tomotani, B. M., H. van der Jeugd, P. Gienapp, I. de la Hera, J. Pilzecker, C. Teichmann, and M. E. Visser (2018). Climate change leads to differential shifts in the timing of annual cycle stages in a migratory bird. Global Change Biology 24:823--835.

\leavevmode\vadjust pre{\hypertarget{ref-twisk2009longitudinal}{}}%
Twisk, J., and F. Rijmen (2009). Longitudinal tobit regression: A new approach to analyze outcome variables with floor or ceiling effects. Journal of Clinical Epidemiology 62:953--958.

\leavevmode\vadjust pre{\hypertarget{ref-underhill1988model}{}}%
Underhill, L. G., and W. Zucchini (1988). \href{https://doi.org/10.1111/j.1474-919x.1988.tb00993.x}{A model for avian primary moult}. Ibis 130:358--372.

\leavevmode\vadjust pre{\hypertarget{ref-underhill1990model}{}}%
Underhill, L. G., W. Zucchini, and R. W. Summers (1990). \href{https://doi.org/10.1111/j.1474-919x.1990.tb01024.x}{A model for avian primary moult-data types based on migration strategies and an example using the redshank \emph{{Tringa} totanus}}. Ibis 132:118--123.

\leavevmode\vadjust pre{\hypertarget{ref-underhill1995relative}{}}%
Underhill, L., and A. Joubert (1995). Relative masses of primary feathers. Ringing \& Migration 16:109--116.

\leavevmode\vadjust pre{\hypertarget{ref-vehtari2017practical}{}}%
Vehtari, A., A. Gelman, and J. Gabry (2017). Practical {Bayesian} model evaluation using leave-one-out cross-validation and WAIC. Statistics and Computing 27:1413--1432.

\leavevmode\vadjust pre{\hypertarget{ref-wang2020extensions}{}}%
Wang, G. (2020). Extensions of the {Underhill-Zucchini} model for avian molt data allowing random duration. University of Washington.

\leavevmode\vadjust pre{\hypertarget{ref-watson1963effect}{}}%
Watson, A. (1963). The effect of climate on the colour changes of mountain hares in {Scotland}. Proceedings of the Zoological Society of London 141:823--835.

\leavevmode\vadjust pre{\hypertarget{ref-watts1996diel}{}}%
Watts, P. (1996). \href{https://doi.org/10.1111/j.1469-7998.1996.tb05494.x}{The diel hauling-out cycle of harbour seals in an open marine environment: Correlates and constraints}. Journal of Zoology 240:175--200.

\leavevmode\vadjust pre{\hypertarget{ref-wolf2000role}{}}%
Wolf, B. O., and G. E. Walsberg (2000). \href{https://doi.org/10.1093/icb/40.4.575}{The role of the plumage in heat transfer processes of birds}. American Zoologist 40:575--584.

\leavevmode\vadjust pre{\hypertarget{ref-wood2017gam}{}}%
Wood, S. N. (2017). Generalized additive models: An introduction with r. 2nd edition. Chapman; Hall/CRC.

\leavevmode\vadjust pre{\hypertarget{ref-zimova2020using}{}}%
Zimova, M., L. S. Barnard, B. M. Davis, A. V. Kumar, D. J. R. Lafferty, and L. S. Mills (2020a). \href{https://doi.org/10.1002/ecs2.3084}{Using remote cameras to measure seasonal molts}. Ecosphere 11:e03084.

\leavevmode\vadjust pre{\hypertarget{ref-zimova2020lack}{}}%
Zimova, M., S. T. Giery, S. Newey, J. J. Nowak, M. Spencer, and L. S. Mills (2020b). Lack of phenological shift leads to increased camouflage mismatch in mountain hares. Proceedings of the Royal Society B 287:20201786.

\leavevmode\vadjust pre{\hypertarget{ref-zimova2022colour}{}}%
Zimova, M., D. Moberg, L. S. Mills, A. J. Dietz, and A. Angerbjörn (2022). Colour moult phenology and camouflage mismatch in polymorphic populations of arctic foxes. Biology Letters 18:20220334.

\end{CSLReferences}

\newpage

\hypertarget{appendix-supplementary-materials}{%
\appendix}

\setcounter{table}{0}  \renewcommand{\thetable}{S\arabic{table}} \setcounter{figure}{0} \renewcommand{\thefigure}{S\arabic{figure}}

\hypertarget{supplementary-materials}{%
\section*{Supplementary materials}\label{supplementary-materials}}
\addcontentsline{toc}{section}{Supplementary materials}

\begin{figure}
\centering
\includegraphics{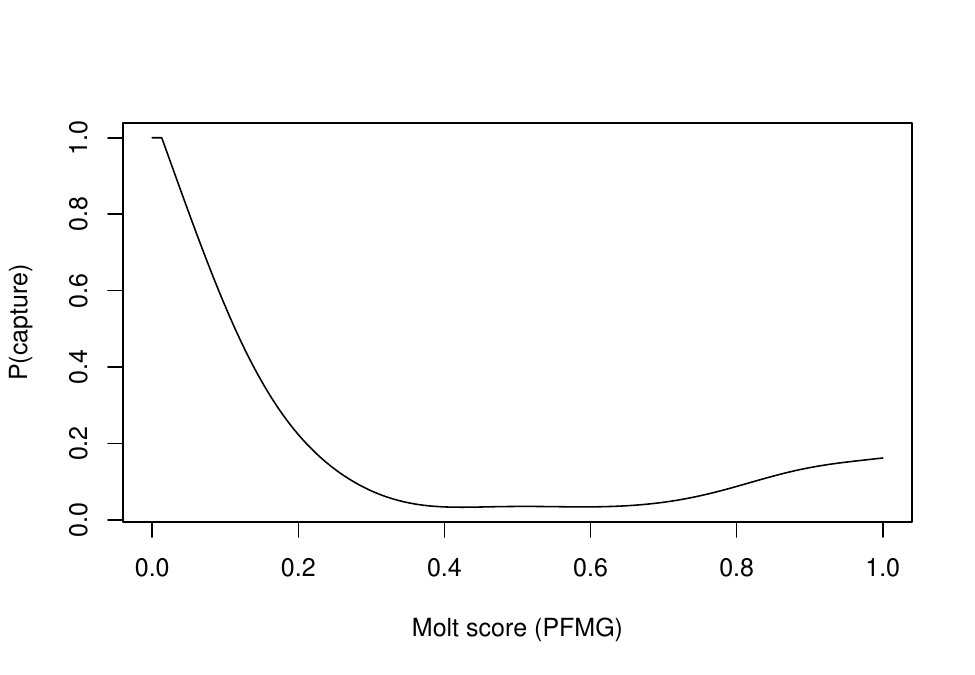}
\caption{\label{fig:siskin-capture-probabilities-figure}All simulation scenarios that generated data with non-constant molt-dependent sampling bias were based on capture probabilities from an empirical data set of Eurasian Sikin \emph{Spinus spinus} captures. Capture probabilities cereased steeply after the onset of moult, and recovered only slightly towards the completion of moult.}
\end{figure}

\hypertarget{sec:software-appendix}{%
\section{\texorpdfstring{Software implementation: The \texttt{moultmcmc} package}{Software implementation: The moultmcmc package}}\label{sec:software-appendix}}

The statistical models in \texttt{moultmcmc} are implemented using the probabilistic programming language Stan (Carpenter et al. 2017, Stan Development Team 2021) which makes use of fast Hamiltonian Monte Carlo samplers.
Individual-level start date intercepts in the recaptures model are calculated using post-sweeping to ensure their identifiability (Ogle and Barber 2020).

The R package is built around pre-compiled Stan models, so that models can be sampled by users without having to compile the Stan code themselves.

Pre-compiled package binaries are available on r-universe (\url{https://pboesu.r-universe.dev}), which allows for fast installation on Windows and MacOS.
Usage examples with code are provided in the package vignettes which can be displayed in a local R session with the command \texttt{vignette(package=\textquotesingle{}moultmcmc\textquotesingle{})} or viewed online at \url{https://pboesu.r-universe.dev/moultmcmc}.

The point-wise log-likelihood for a model can be obtained from the fitting functions.
This can be used for cross-validation methods such as leave-one-out cross-validation (LOO-CV, Vehtari et al. 2017), which are useful for model comparison, selection, or averaging.

\hypertarget{software-installation}{%
\subsection{Software installation}\label{software-installation}}

The easiest and quickest way of installing \texttt{moultmcmc} is to install the package from R-universe using the following code:

\begin{Shaded}
\begin{Highlighting}[]
\FunctionTok{install.packages}\NormalTok{(}\StringTok{"moultmcmc"}\NormalTok{, }\AttributeTok{repos =} \StringTok{"https://pboesu.r{-}universe.dev"}\NormalTok{)}
\end{Highlighting}
\end{Shaded}

On MacOS and Windows systems this will make use of pre-compiled binaries, which means the models can be installed without a C++ compiler toolchain. On Linux this will install the package from a source tarball. Because of the way the Stan models are currently structured, compilation from source can be a lengthy process (15-45 minutes), depending on system setup and compiler toolchain (where possible it is strongly recommended to compile using multiple threads).

To install \texttt{moultmcmc} directly from the github source use the following code. This requires a working C++ compiler and a working installation of \texttt{rstan}:

\begin{Shaded}
\begin{Highlighting}[]
\CommentTok{\#not generally recommended for Windows or MacOS users}
\FunctionTok{install.packages}\NormalTok{(}\StringTok{"remotes"}\NormalTok{)}
\NormalTok{remotes}\SpecialCharTok{::}\FunctionTok{install\_github}\NormalTok{(}\StringTok{"pboesu/moultmcmc"}\NormalTok{)}
\end{Highlighting}
\end{Shaded}

\hypertarget{sec:std-model-appendix}{%
\section{Comparison of parameter estimates for standard Underhill-Zucchini models}\label{sec:std-model-appendix}}

To compare estimation methods for standard Underhill-Zucchini models we fitted type 1-5 molt models to primary feather data from 164 Sanderlings (\emph{Calidris alba}) trapped on 11 days in the southwestern Cape, South Africa, between October 1978 and April 1979 (Underhill and Zucchini 1988).
Molt parameters were estimated using both the maximum likelihood approach implement in R package \emph{moult} (Erni et al. 2013) and the Bayesian approach implemented in R package \emph{moultmcmc} (Boersch-Supan et al. 2023a).
Both estimation methods produced comparable results.

\begin{figure}
\includegraphics[width=29.17in]{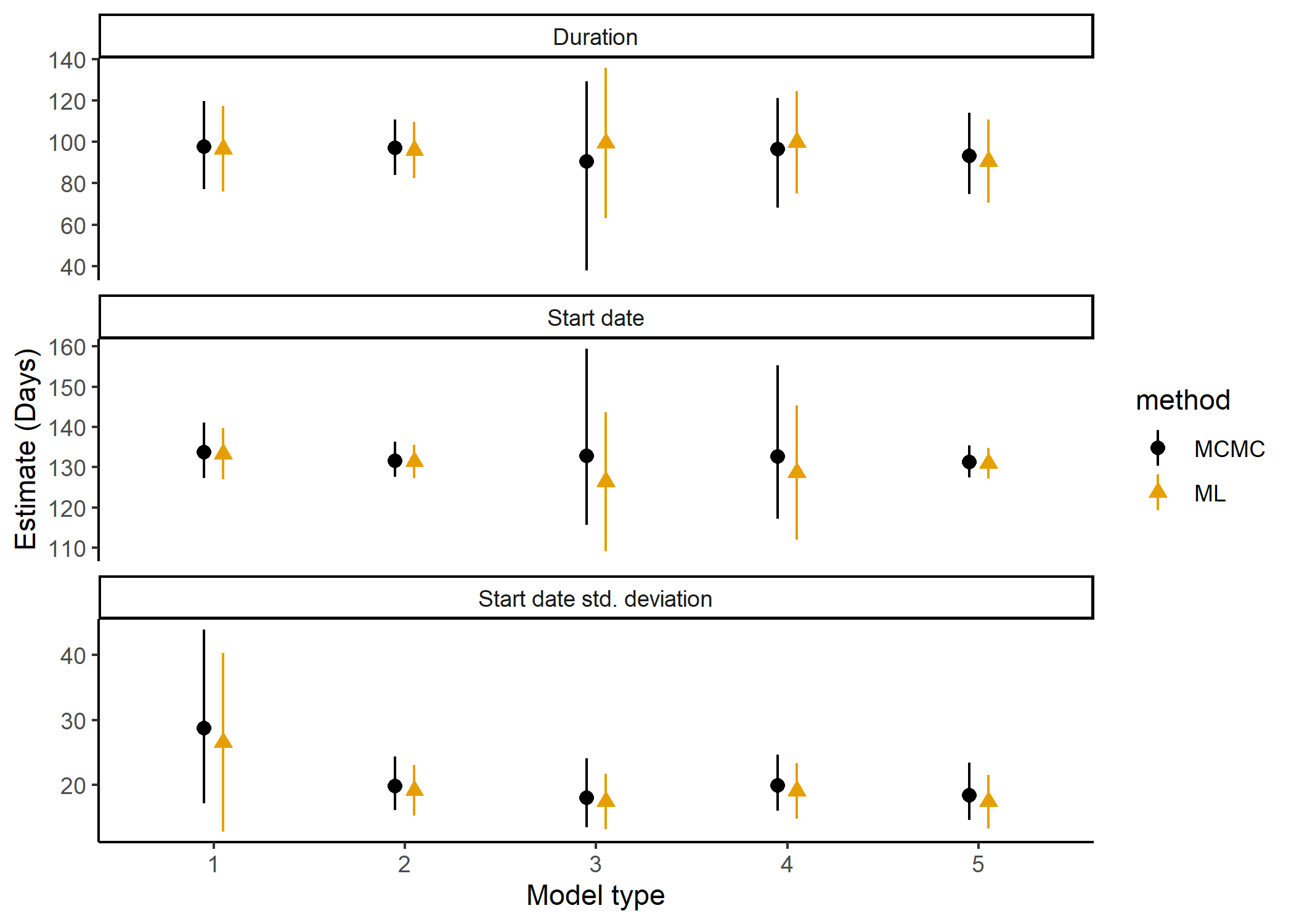} \caption{Maximum likelihood (ML) and Bayesian (MCMC) estimates for molt parameters are in close agreement for the standard Underhill-Zucchini models.}\label{fig:supp-fig-mcmc-ml}
\end{figure}

\hypertarget{sec:t3-appendix}{%
\section{Parameter uncertainty in the Type 3 model}\label{sec:t3-appendix}}

To assess parameter identifiability in the Type 3 model we simulated 200 datasets with random combinations of molt durations between 35 and 200 days, and \(\sigma/\tau\) values between 0.01 and 0.6 - spanning a wide range of biologically plausible combinations (Ginn and Melville 1983, Oschadleus 2005).
Molt parameters were then estimated from the simulated data using the T1, T2, and T3 models, and the obtained parameter estimates and their estimated precision were compared.

\begin{figure}
\includegraphics[width=25in]{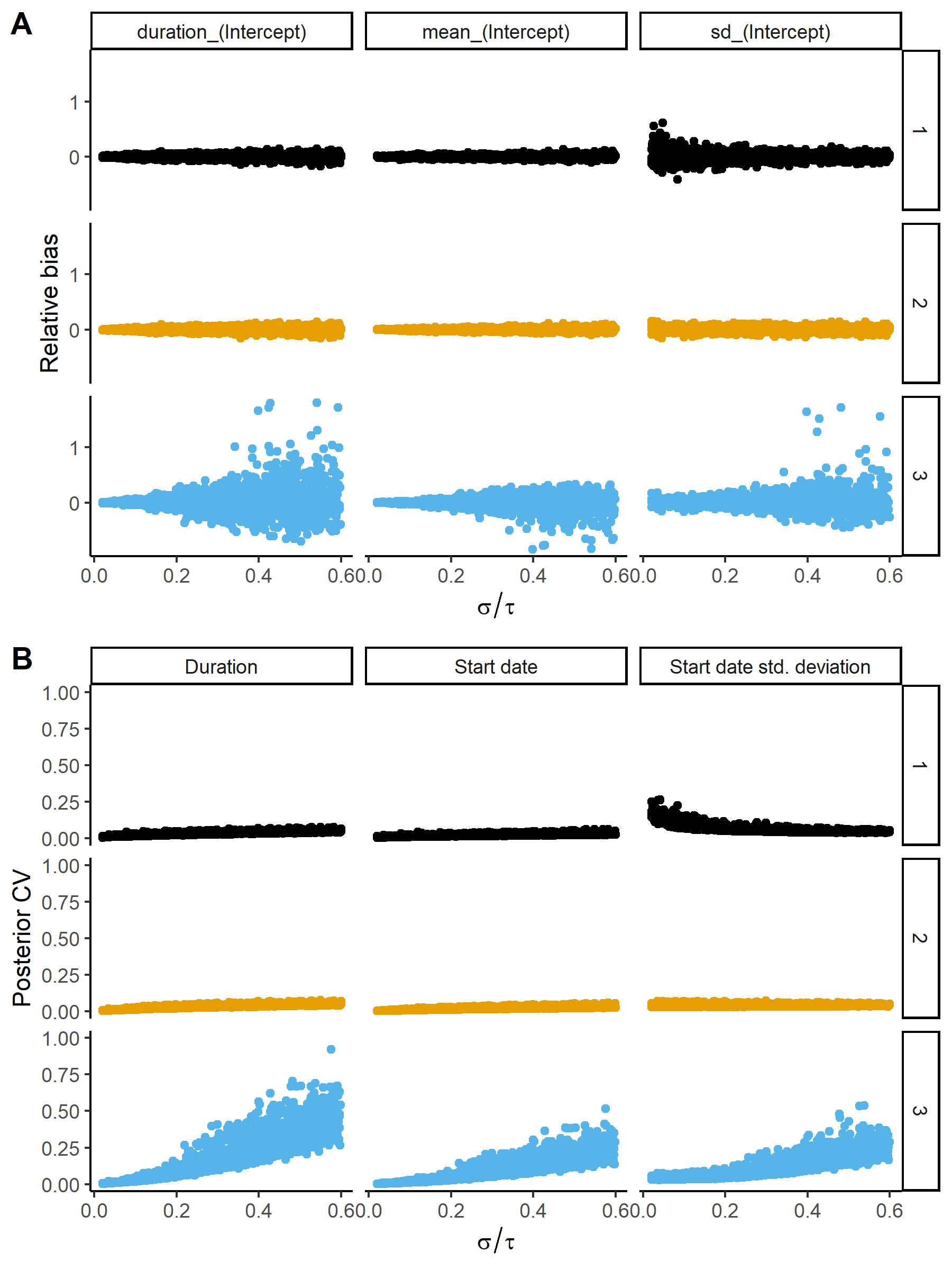} \caption{When the standard deviation of the start date of molt is large relative to molt duration, the estimates  of molt parameters are strongly correlated in the Type 3 model. Although unbiased across simulations, accuracy (A) and precision (B) of estimates for individual datasets suffer as the ratio $\sigma/\tau$ increases. Points represent parameter estimates for 200 simulated datasets.}\label{fig:supp-fig-t3-identifiability}
\end{figure}
\clearpage

\hypertarget{sec:t1r-appendix}{%
\section{The type 1 recapture model for high-frequency recapture data}\label{sec:t1r-appendix}}

Mark-recapture datasets based on physical trapping methods are often characterised by relatively low numbers of recaptures per individual within a given season, e.g.~because trapping effort is limited by logistical constraints, trapping efficiency is low, and/or trap shyness is high.
If molt data is recorded as a categorical variable, low numbers of recaptures severely limit the ability to draw individual-level inferences and/or exploit the recaptures model for improved population-level inferences.
Much higher rates of within-season recaptures can be achieved using non-invasive recapture methods such as camera traps (e.g., Zimova et al. 2020a).
We here demonstrate the performance of the type 1 recaptures model for such high-frequency recapture data using a set of camera trap observations of Arctic Foxes (\emph{Vulpes lagopus}; Zimova et al. (2022)).
We used a subset of 1444 observations between year day 1 and year day 221 (Aug 9) to estimate the spring molt phenology of 65 individuals of which 61 were recaptured at least once in at least one of two years.
Individuals were observed on average 19 times per year (SD 13, range 1-54).

\begin{figure}
\includegraphics[width=6.5in]{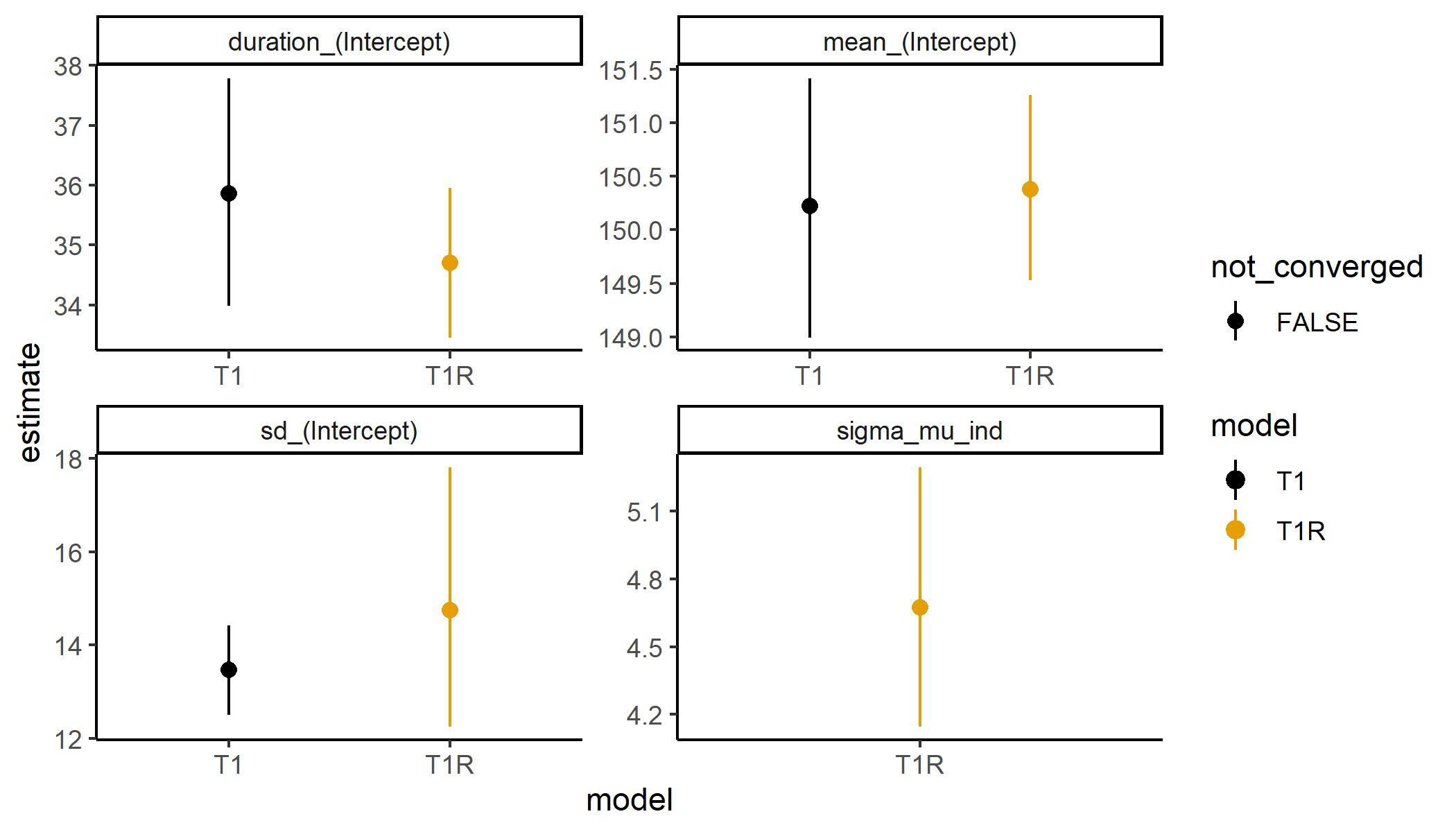} \caption{Parameter estimates from T1 and T1R models fitted to a camera trap dataset of Arctic Fox molt observations (Zimova et al. 2022). At high individual recapture rates the T1R model provides more precise estimates of start date and duration than the standard T1 model ignoring repeated measures.  The less precise estimate of the population standard deviation of the molt start date in the T1R model reflects the lower (true) effective sample size of 75 individual-year combinations.}\label{fig:supp-fig-t1r}
\end{figure}

\begin{figure}
\includegraphics[width=6.5in]{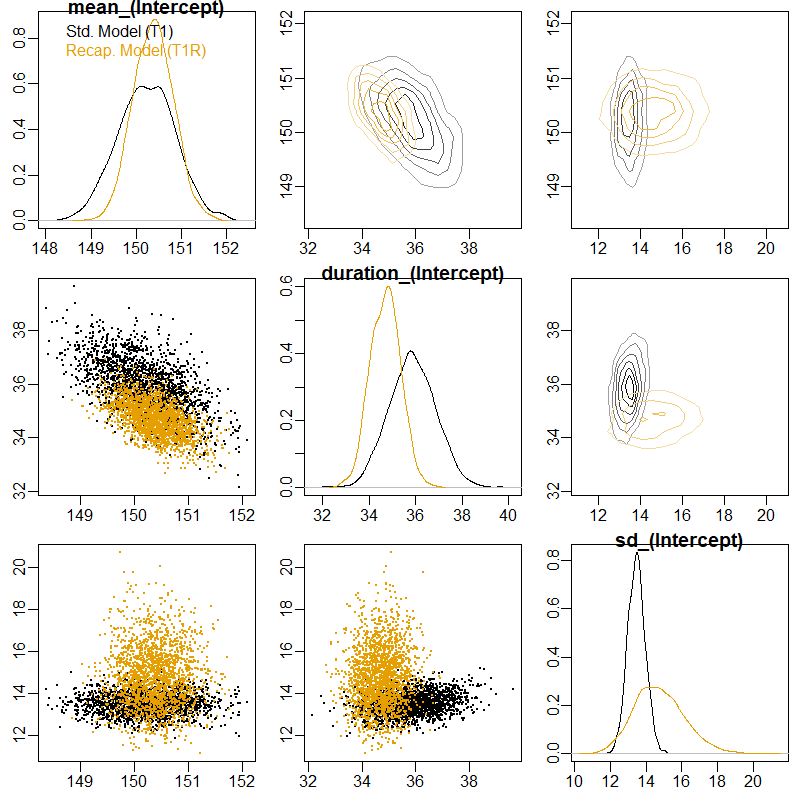} \caption{Uni- and bivariate distributions of posterior draws from T1 and T1R models fitted to a camera trap dataset of Arctic Fox molt observations (Zimova et al. 2022). Recaptures information taken into account by the T1R model reduces the posterior correlation between start date and duration estimates.  The less precise estimate of the population standard deviation of the molt start date in the T1R model reflects the lower (true) effective sample size.}\label{fig:supp-fig-t1r-compairs}
\end{figure}

Both models estimated a molt start date around year day 150 (May 30) and a molt duration of 36 days (Fig. \ref{fig:supp-fig-t1r}).
The recaptures model provided more estimates of start date and duration, but a less precise estimate of the population standard deviation of the molt start date.
This is expected given the high degree of pseudo-replication in the standard T1 model.
Pairwise plots of posterior parameter draws show that the increase in precision is a result of a reduction in the posterior correlation of start date and duration estimates (Fig. \ref{fig:supp-fig-t1r-compairs}).

\clearpage

\hypertarget{worked-example---fitting-molt-models-to-real-world-data}{%
\section{Worked example - fitting molt models to real-world data}\label{worked-example---fitting-molt-models-to-real-world-data}}

The following code walks through fitting Bayesian molt models to real-world observations from American Redstart (Mumme et al. 2021b).
The example assumes that the data have been downloaded from the corresponding Dryad repository (\url{https://datadryad.org/stash/dataset/doi:10.5061/dryad.63xsj3v0x}) and extracted to a suitable directory.
The data are then read in and the primary scores are converted to percentage feather mass grown (PFMG) using generic feather mass indices from Bonnevie (2010b).

\begin{Shaded}
\begin{Highlighting}[]
\FunctionTok{library}\NormalTok{(dplyr)}
\FunctionTok{library}\NormalTok{(ggplot2)}
\FunctionTok{library}\NormalTok{(moult)}
\FunctionTok{library}\NormalTok{(moultmcmc)}

\CommentTok{\#generic feather mass indices for a passerine with nine primaries (Bonnevie 2010b)}
\NormalTok{fmi\_passer10 }\OtherTok{\textless{}{-}} \FunctionTok{c}\NormalTok{(}\FloatTok{8.0}\NormalTok{, }\FloatTok{8.6}\NormalTok{, }\FloatTok{9.2}\NormalTok{, }\FloatTok{10.0}\NormalTok{, }\FloatTok{11.4}\NormalTok{, }\FloatTok{12.4}\NormalTok{, }\FloatTok{13.0}\NormalTok{, }\FloatTok{13.5}\NormalTok{, }\FloatTok{14.0}\NormalTok{) }

\CommentTok{\#read AMRE data}
\CommentTok{\# replace path with relevant path to the downloaded data}
\NormalTok{amre\_all }\OtherTok{\textless{}{-}}\NormalTok{ readr}\SpecialCharTok{::}\FunctionTok{read\_csv}\NormalTok{(}\StringTok{\textquotesingle{}../data/digitized/Mumme2021/AllRawData.csv\textquotesingle{}}\NormalTok{) }\SpecialCharTok{\%\textgreater{}\%} 
  \FunctionTok{filter}\NormalTok{(Species }\SpecialCharTok{==} \StringTok{\textquotesingle{}AMRE\textquotesingle{}}\NormalTok{) }\SpecialCharTok{\%\textgreater{}\%} \CommentTok{\# discard all other species}
\NormalTok{  tidyr}\SpecialCharTok{::}\FunctionTok{unite}\NormalTok{(}\AttributeTok{col =} \StringTok{\textquotesingle{}moult\_str\textquotesingle{}}\NormalTok{, P1}\SpecialCharTok{:}\NormalTok{P9, }\AttributeTok{sep =} \StringTok{\textquotesingle{}\textquotesingle{}}\NormalTok{, }\AttributeTok{remove =}\NormalTok{ F) }\SpecialCharTok{\%\textgreater{}\%} \CommentTok{\# collate primary scores}
  \FunctionTok{mutate}\NormalTok{(}\AttributeTok{pfmg =}\NormalTok{ moult}\SpecialCharTok{::}\FunctionTok{ms2pfmg}\NormalTok{(moult\_str, fmi\_passer10), }\CommentTok{\# convert to PFMG}
         \AttributeTok{ring\_yr =} \FunctionTok{as.factor}\NormalTok{(}\FunctionTok{paste}\NormalTok{(Suffix, YEAR, }\AttributeTok{sep =} \StringTok{"\_"}\NormalTok{))) }
\CommentTok{\# create identifier for within{-}year recaptures of individuals}
\end{Highlighting}
\end{Shaded}

We can then visualise the data, highlighting the undersampling of birds at intermediate molt stages.

\begin{Shaded}
\begin{Highlighting}[]
\NormalTok{amre\_all }\SpecialCharTok{\%\textgreater{}\%} \FunctionTok{ggplot}\NormalTok{(}\FunctionTok{aes}\NormalTok{(}\AttributeTok{x =}\NormalTok{ pfmg)) }\SpecialCharTok{+}
  \FunctionTok{geom\_histogram}\NormalTok{(}\AttributeTok{binwidth =} \FloatTok{0.1}\NormalTok{) }\SpecialCharTok{+}
  \FunctionTok{theme\_classic}\NormalTok{() }\SpecialCharTok{+}
  \FunctionTok{ylab}\NormalTok{(}\StringTok{\textquotesingle{}Number of records\textquotesingle{}}\NormalTok{) }\SpecialCharTok{+}
  \FunctionTok{xlab}\NormalTok{(}\StringTok{\textquotesingle{}PFMG\textquotesingle{}}\NormalTok{)}
\end{Highlighting}
\end{Shaded}

\includegraphics{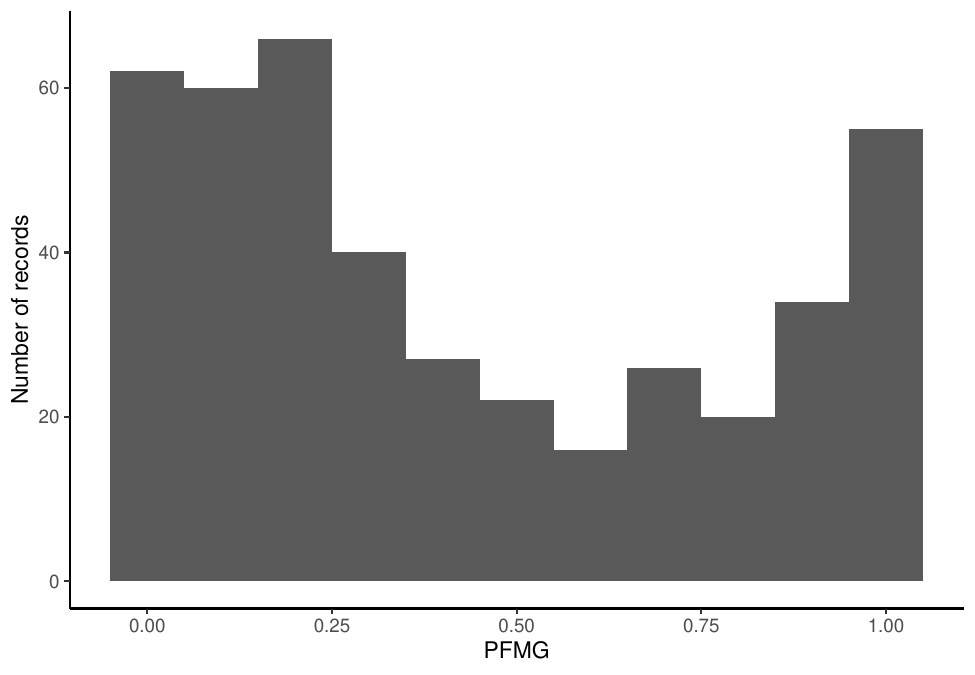}

We can also visulaise the seasonal progression of molt progress and highlight the recaptures.

\begin{Shaded}
\begin{Highlighting}[]
\NormalTok{amre\_all }\SpecialCharTok{\%\textgreater{}\%} \FunctionTok{ggplot}\NormalTok{(}\FunctionTok{aes}\NormalTok{(}\AttributeTok{x =}\NormalTok{ Julian, }\AttributeTok{y =}\NormalTok{ pfmg, }\AttributeTok{group =}\NormalTok{ ring\_yr)) }\SpecialCharTok{+}
  \FunctionTok{geom\_line}\NormalTok{(}\AttributeTok{alpha=}\FloatTok{0.3}\NormalTok{) }\SpecialCharTok{+}
  \FunctionTok{geom\_point}\NormalTok{(}\AttributeTok{alpha =} \FloatTok{0.3}\NormalTok{) }\SpecialCharTok{+}
  \FunctionTok{theme\_classic}\NormalTok{()}\SpecialCharTok{+}
  \FunctionTok{xlab}\NormalTok{(}\StringTok{\textquotesingle{}Days since Jan 01\textquotesingle{}}\NormalTok{) }\SpecialCharTok{+}
  \FunctionTok{ylab}\NormalTok{(}\StringTok{\textquotesingle{}PFMG\textquotesingle{}}\NormalTok{)}
\end{Highlighting}
\end{Shaded}

\includegraphics{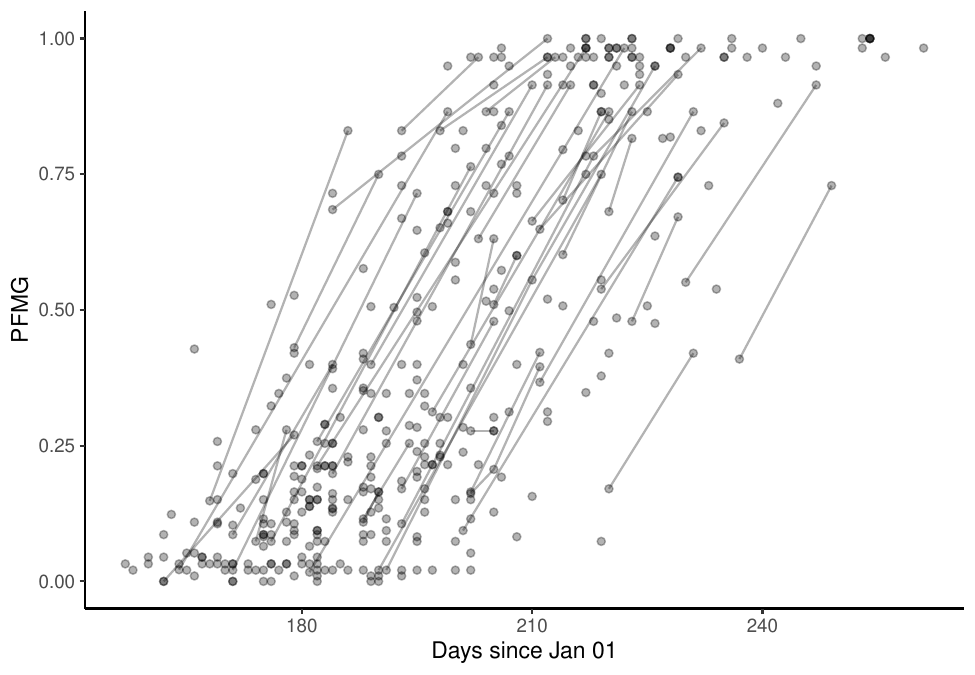}

We then fit both a standard type 3 Underhill-Zucchini molt phenology model, and an extended type 3 recaptures model to demonstrate how both models perform in the face of the molt-dependent sampling bias.
The standard T3 model fits in a few seconds, therecaptures model is computationally much more intensive and takes approximately 5-10 minutes to fit.

\begin{Shaded}
\begin{Highlighting}[]
\NormalTok{amre\_t3 }\OtherTok{\textless{}{-}} \FunctionTok{moultmcmc}\NormalTok{(}\AttributeTok{moult\_column =} \StringTok{\textquotesingle{}pfmg\textquotesingle{}}\NormalTok{,}
                      \AttributeTok{date\_column =} \StringTok{\textquotesingle{}Julian\textquotesingle{}}\NormalTok{,}
                      \AttributeTok{data =}\NormalTok{ amre\_all,}
                      \AttributeTok{type =} \DecValTok{3}\NormalTok{,}
                     \AttributeTok{iter =} \DecValTok{1000}\NormalTok{,}
                     \AttributeTok{chains =} \DecValTok{2}\NormalTok{,}
                     \AttributeTok{cores =} \DecValTok{2}\NormalTok{)}\CommentTok{\# The standard model fits in c. 5 seconds}
\NormalTok{amre\_t3r }\OtherTok{\textless{}{-}} \FunctionTok{moultmcmc}\NormalTok{(}\AttributeTok{moult\_column =} \StringTok{\textquotesingle{}pfmg\textquotesingle{}}\NormalTok{,}
                      \AttributeTok{date\_column =} \StringTok{\textquotesingle{}Julian\textquotesingle{}}\NormalTok{,}
                 \AttributeTok{id\_column =} \StringTok{"ring\_yr"}\NormalTok{,}
                      \AttributeTok{data =}\NormalTok{ amre\_all,}
                      \AttributeTok{type =} \DecValTok{3}\NormalTok{,}
                 \AttributeTok{iter =} \DecValTok{1000}\NormalTok{,}
                 \AttributeTok{chains =} \DecValTok{2}\NormalTok{,}
                 \AttributeTok{cores =} \DecValTok{2}\NormalTok{)}\CommentTok{\#the recaptures model fits in c. 6 minutes}
\end{Highlighting}
\end{Shaded}

We can then plot the model fits.

\begin{Shaded}
\begin{Highlighting}[]
\FunctionTok{moult\_plot}\NormalTok{(amre\_t3) }\SpecialCharTok{+} \FunctionTok{ggtitle}\NormalTok{(}\StringTok{\textquotesingle{}T3 model\textquotesingle{}}\NormalTok{)}
\end{Highlighting}
\end{Shaded}

\includegraphics{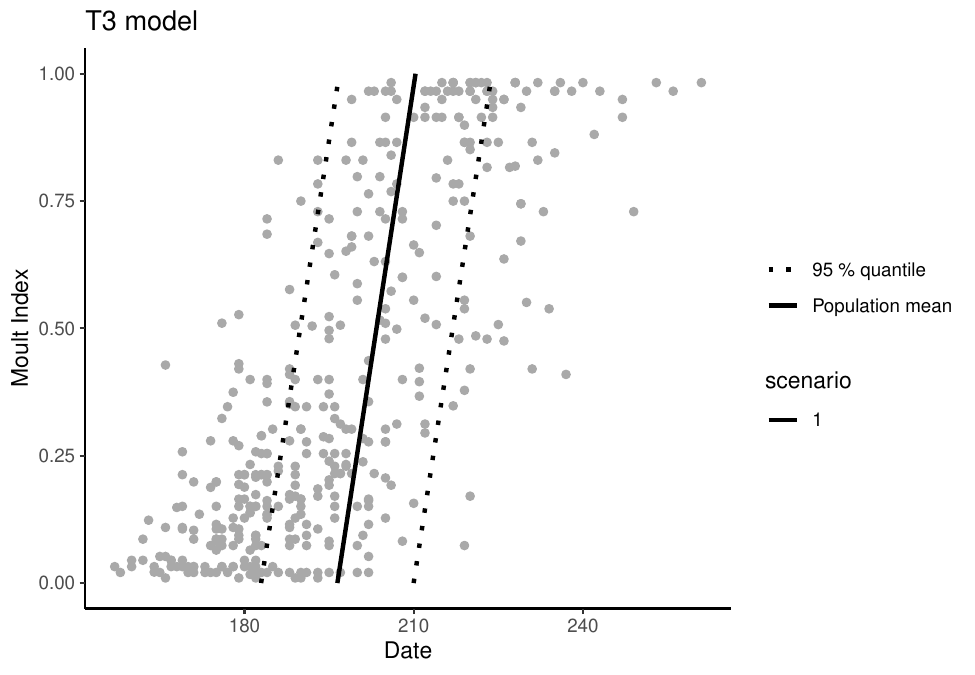}

\begin{Shaded}
\begin{Highlighting}[]
\FunctionTok{moult\_plot}\NormalTok{(amre\_t3r) }\SpecialCharTok{+} \FunctionTok{ggtitle}\NormalTok{(}\StringTok{\textquotesingle{}T3R model\textquotesingle{}}\NormalTok{)}
\end{Highlighting}
\end{Shaded}

\includegraphics{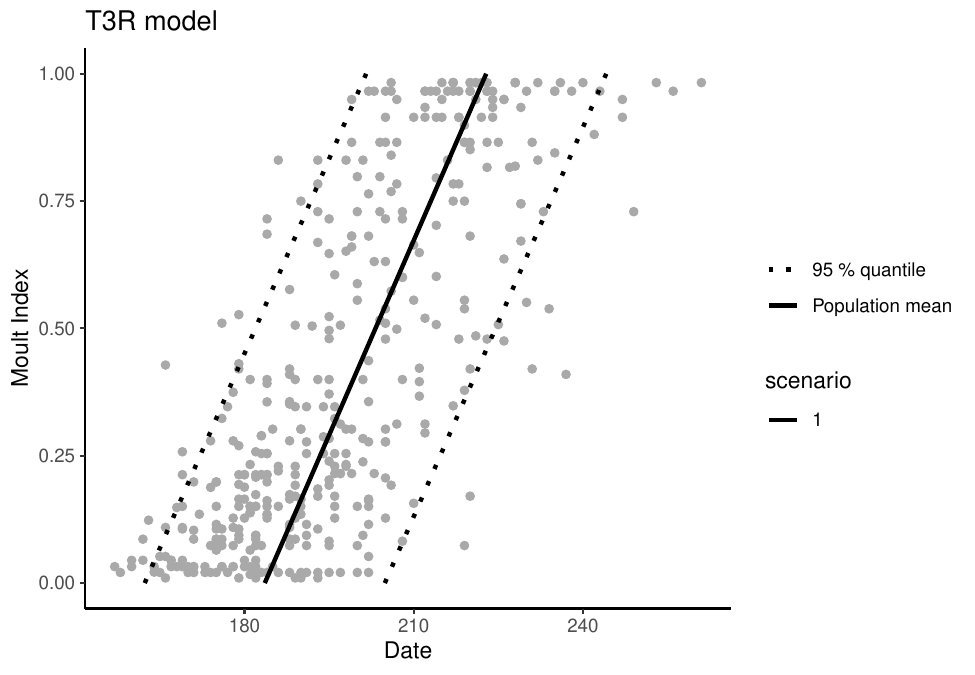}

The model plots provide a relatively straightforward visual check of biased estimates. Poor coverage of the observations by the 95\% molt interval shows that the T3 model fits the data poorly, whereas the molt interval for the T3R fit encompasses the observations well.

Additional usage examples for the moultmcmc package can be found in the package vignettes at \url{https://pboesu.r-universe.dev/moultmcmc}.
Additional case studies with real-world data and code examples can be found in Boersch-Supan et al. (2023b).
\clearpage

\hypertarget{sec:likelihoods-appendix}{%
\section{Molt model likelihoods}\label{sec:likelihoods-appendix}}

\hypertarget{type-1}{%
\subsection*{Type 1}\label{type-1}}
\addcontentsline{toc}{subsection}{Type 1}

Type 1 data consist of observations of categorical molt state (pre-molt, active molt, post-molt) and sampling is representative in all three categories. The likelihood of these observations is
\[
\mathcal{L}(\boldsymbol\theta,t,u,v) = \prod_{i=1}^IP(t_i)\prod_{j = 1}^JQ(u_j)\prod_{k=1}^KR(v_k).
\]

\hypertarget{lumped-type-1}{%
\subsection*{Lumped Type 1}\label{lumped-type-1}}
\addcontentsline{toc}{subsection}{Lumped Type 1}

Lumped type 1 data consist of observations of individuals in all three molt states (pre-molt, active molt, post-molt), but where the pre-molt and post-molt states cannot be distinguished from each other, yielding \(I\) observations of dates \(t_i\) on which non-molting individuals were observed. Sampling is representative for all three categories. The likelihood of these observations is
\[
\mathcal{L}(\boldsymbol\theta,t,u,y) = \prod_{i=1}^IPR(t_i)\prod_{j = 1}^JQ(u_j),
\]
where \(PR(t_i)=P(t_i)+R(t_i).\)

\hypertarget{type-2}{%
\subsection*{Type 2}\label{type-2}}
\addcontentsline{toc}{subsection}{Type 2}

Type 2 data consist of observations of individuals in all three molt states (pre-molt, active molt, post-molt). Sampling is representative for all three categories, and for actively molting individuals a sufficiently linear molt index \(y\) (e.g.~percent feather mass grown) is known. The likelihood of these observations is
\[
\mathcal{L}(\boldsymbol\theta,t,u,y,v) = \prod_{i=1}^IP(t_i)\prod_{j = 1}^Jq(u_j,y_j)\prod_{k=1}^KR(v_k),
\]
where \(q(u_j,y_j) = \tau f_T(u_j-y_j\tau).\)

\hypertarget{lumped-type-2}{%
\subsection*{Lumped Type 2}\label{lumped-type-2}}
\addcontentsline{toc}{subsection}{Lumped Type 2}

Lumped type 2 data consist of observations of individuals in all three molt states (pre-molt, active molt, post-molt), but where the pre-molt and post-molt states cannot be distinguished from each other, yielding \(I\) observations of dates \(t_i\) on which non-molting individuals were observed. Sampling is representative for all three categories, and for actively molting individuals a sufficiently linear molt index \(y\) (e.g.~percent feather mass grown) is known. The likelihood of these observations is
\[
\mathcal{L}(\boldsymbol\theta,t,u,y) = \prod_{i=1}^IPR(t_i)\prod_{j = 1}^Jq(u_j,y_j),
\]
where \(q(u_j,y_j) = \tau f_T(u_j-y_j\tau)\) and \(PR(t_i)=P(t_i)+R(t_i).\)

\hypertarget{type-3}{%
\subsection*{Type 3}\label{type-3}}
\addcontentsline{toc}{subsection}{Type 3}

Type 3 data consist of observations of actively molting individuals only, and a sufficiently linear molt index \(y\) (e.g.~percent feather mass grown) is known for each individual. The likelihood of these observations is
\[
\mathcal{L}(\boldsymbol\theta,u,y) = \prod_{j = 1}^J\frac{q(u_j,y_j)}{Q(u_j)}.
\]

\hypertarget{type-4}{%
\subsection*{Type 4}\label{type-4}}
\addcontentsline{toc}{subsection}{Type 4}

Type 4 data consist of observations of individuals in active molt and post-molt only. Sampling is representative for these two categories, and for actively molting individuals a sufficiently linear molt index \(y\) (e.g.~percent feather mass grown) is known. The likelihood of these observations is
\[
\mathcal{L}(\boldsymbol\theta,u,y,v) = \prod_{j = 1}^J\frac{q(u_j,y_j)}{1-P(u_j)}\prod_{k=1}^K\frac{R(v_k)}{1-P(v_k)}.
\]

\hypertarget{type-5}{%
\subsection*{Type 5}\label{type-5}}
\addcontentsline{toc}{subsection}{Type 5}

Type 5 data consist of observations of individuals in pre-molt and active molt. Sampling is representative for these two categories, and for actively molting individuals a sufficiently linear molt index \(y\) (e.g.~percent feather mass grown) is known. The likelihood of these observations is
\[
\mathcal{L}(\boldsymbol\theta,t,u,y) = \prod_{i=1}^I\frac{P(t_i)}{1-R(t_i)}\prod_{j = 1}^J\frac{q(u_j,y_j)}{1-R(u_j)}.
\]

\hypertarget{type-1-2}{%
\subsection*{Type 1 + 2}\label{type-1-2}}
\addcontentsline{toc}{subsection}{Type 1 + 2}

As outlined in Underhill and Zucchini (1988) estimates can also be derived from mixtures of data types. Type 1 + 2 data consist of observations of individuals in all three molt states (pre-molt, active molt, post-molt). Sampling is representative for all three categories, but a sufficiently linear molt index \(y\) (e.g.~percent feather mass grown) is known only for some of the actively molting individuals. This means the sample consist of \(I\) pre-molt individuals, \(J\) individuals in active molt with known indices, \(L\) individuals in active molt without known indices but known capture dates \(u'=u'_l,\ldots,u'_L\), and \(K\) post-molt individuals.
The likelihood of these observations is
\[
\mathcal{L}(\boldsymbol\theta,t,u,y,u',v) = \prod_{i=1}^IP(t_i)\prod_{j = 1}^Jq(u_j,y_j)\prod_{l = 1}^{L}Q(u'_{l})\prod_{k=1}^KR(v_k),
\]

\hypertarget{recapture-models}{%
\subsection*{Recapture models}\label{recapture-models}}
\addcontentsline{toc}{subsection}{Recapture models}

\emph{moultmcmc} currently implements a recaptures model which allows for heterogeneity in start dates \(\mu\) but assumes a common molt duration \(\tau\).
When repeat observations are available an individual's start date \(\mu_n\) then becomes

\begin{equation}
\mu_n = \mu_0 + \mu'_n + \mathbf{x}_\mu\boldsymbol{\beta}_\mu 
\end{equation}

where \(\boldsymbol{x}_\mu\) is a row vector containing the values of individual-specific predictors (in the same format as \(\boldsymbol{X}_\mu\)), and \(\mu'_n\) is an individual-level random effect intercept

\begin{equation}
\mu'_n \sim \mathrm{Normal}(0,\sigma_n) 
\end{equation}
where \(\sigma_n\) is the individual-specific standard deviation.
We can then exploit the linearity assumption and treat observed molt scores \(y_{ni}\) on dates \(u_{ni}\) as
\begin{equation}
 u_{ni} \sim \mathrm{Normal}(\mu_0 + \mu'_n + \tau * y_{ni}, \sigma_\tau) 
\end{equation}
where \(\sigma_\tau\) captures any unmodelled variance in \(\tau\) as well as any measurement error in \(y\).

The likelihood for the Type 3-like model for a sample of \(J\) individuals in active molt without repeated observations, and \(N\) individuals in active molt with a total of \(M\) repeated observations \(u'=u'_m,\ldots,u'_M\) and \(y'=y'_m,\ldots,y'_M\) then is

\[
\mathcal{L}(\boldsymbol\theta,u,y,u',y') = \prod_{j = 1}^J\frac{q(u_j,y_j)}{Q(u_j)}\prod_{m = 1}^{M}f(u'_m,y'_m)\prod_{n=1}^N\phi(\mu'_n|0,\sigma_n),
\]
where \(f(u'_m,y'_m)\) follows from above.

\newpage

\hypertarget{sec:priors-appendix}{%
\section{Model priors}\label{sec:priors-appendix}}

By default flat priors are used for \(\mu_0\) and \(\tau_0\) and a vaguely informative half-normal prior on \(\ln(\sigma_0)\)

\(\mu_0 \sim \mathrm{Uniform(0,366)}\)\\
\(\tau_0 \sim \mathrm{Uniform(0,366)}\)\\
\(\ln(\sigma_0) \sim \mathrm{Normal(0,5)}\)

In some cases the models sample poorly with these priors, and better convergence can be achieved using vaguely informative truncated normal priors for \(\mu_0\) and \(\tau_0\):

\(\mu_0 \sim \mathrm{TruncNormal(150,50,0,366)}\)\\
\(\tau_0 \sim \mathrm{TruncNormal(100,30,0,366)}\)

These priors work well for data from passerines in seasonal environments, i.e.~when the sampling occasion data is encoded as days from the middle of the non-breeding season.

For any additional regression coefficients an improper flat prior is used as a default.

\end{document}